%% file: main.tex
\documentclass[sigconf]{acmart}
\usepackage{footnote}
\makesavenoteenv{tabular}
\makesavenoteenv{table}

\PassOptionsToPackage{bookmarks=true,breaklinks=true,letterpaper=true,colorlinks,linkcolor=black,citecolor=black,urlcolor=black}{hyperref}

\copyrightyear{2021} 
\acmYear{2021} 
\setcopyright{rightsretained} 
\acmConference[MICRO '21]{MICRO'21: 54th Annual IEEE/ACM International Symposium on Microarchitecture}{October 18--22, 2021}{Virtual Event, Greece}
\acmBooktitle{MICRO'21: 54th Annual IEEE/ACM International Symposium on Microarchitecture (MICRO '21), October 18--22, 2021, Virtual Event, Greece}\acmDOI{10.1145/3466752.3480069}
\acmISBN{978-1-4503-8557-2/21/10}

\DeclareMathAlphabet{\mathcal}{OMS}{cmsy}{m}{n}
\usepackage[tt=false]{libertine}

\usepackage{array}
\usepackage{eso-pic}

\usepackage{fancyhdr}
\usepackage{datetime}
\usepackage[bookmarks=true,breaklinks=true,letterpaper=true,colorlinks,linkcolor=black,citecolor=black,urlcolor=black]{hyperref}
\usepackage{multirow}
\usepackage{multicol}
\usepackage{float}
\usepackage[shortlabels]{enumitem}
\usepackage[nomessages]{fp}
\usepackage{dblfloatfix}    
\usepackage{bm}
\usepackage{tikz}
\usepackage{atbegshi}
\usepackage{xcolor}
\usepackage[font={small,bf}]{subcaption}
\usepackage[linesnumbered,ruled]{algorithm2e}
\usepackage{balance}

\def\BibTeX{{\rm B\kern-.05em{\sc i\kern-.025em b}\kern-.08em
    T\kern-.1667em\lower.7ex\hbox{E}\kern-.125emX}}

\usepackage[font={small,bf}, figurename=Fig.]{caption}
\setlist[itemize]{leftmargin=*}

\usepackage{listings}
\usepackage{amsmath}

\usepackage{siunitx} 
\usepackage{glossaries}
\usepackage{makecell} 
\newcommand{\todo}[1]{TODO HERE}

 \setlist[itemize]{noitemsep,itemsep=0pt,parsep=0pt,topsep=0pt,partopsep=0pt,leftmargin=1.5em}
 \setlist[enumerate]{noitemsep,itemsep=0pt,parsep=0pt,topsep=0pt,partopsep=0pt,leftmargin=1.5em}

\usepackage{titlesec}
\titlespacing\section{0pt}{3pt plus 1pt minus 1pt}{2pt plus 1pt minus 1pt}
\titlespacing\subsection{0pt}{3pt plus 1pt minus 1pt}{2pt plus 1pt minus 1pt}
\titlespacing\subsubsection{0pt}{3pt plus 1pt minus 1pt}{2pt plus 1pt minus 1pt}

\input{macros}
 \renewcommand{\agyl}[1]{\agycomment{#1}}
 \renewcommand{\oml}[1]{\onurcomment{#1}}
 \renewcommand{\hluol}[1]{\hluocomment{#1}}

\makeatletter
\newcommand*{\textoverline}[1]{$\overline{\raisebox{0pt}[0.85\height]{#1}}\m@th$}
\makeatother

\begin{document}

\title[A Deeper Look into RowHammer’s Sensitivities: Experimental Analysis of Real DRAM Chips\\and Implications on Future Attacks and Defenses]{A Deeper Look into RowHammer’s Sensitivities:\\Experimental Analysis of Real DRAM Chips\\and Implications on Future Attacks and Defenses}

\newcommand{\ethz}{{\large$^\dagger$}}
\newcommand{\cmu}{{\large$^\ddagger$}}
\newcommand{\scomma}{{\large$^,$}}

\newcommand{\affilTOBB}[0]{\textsuperscript{$\ddagger$}}
\newcommand{\coauthor}[0]{\textsuperscript{$\ast$}}

\settopmatter{authorsperrow=5}

\author{Lois Orosa}
\authornote{These authors equally contributed to this work.}
\affiliation{%
  \institution{ETH Z{\"u}rich}
  \country{}
}

\author{A. Giray Ya{\u{g}}l{\i}k{\c{c}}{\i}}
\authornotemark[1]
\affiliation{%
  \institution{ETH Z{\"u}rich}
  \country{}
}

\author{Haocong Luo}
\affiliation{%
  \institution{ETH Z{\"u}rich}
  \country{}
}
\author{Ataberk Olgun}
\affiliation{%
  \institution{ETH~Z{\"u}rich,~TOBB~ET{\"U}}
  \country{}
}
\author{Jisung Park}
\affiliation{%
  \institution{ETH Z{\"u}rich}
  \country{}
}
\author{Hasan Hassan}
\affiliation{%
  \institution{ETH Z{\"u}rich}
  \country{}
}
\author{Minesh Patel}
\affiliation{%
  \institution{ETH Z{\"u}rich}
  \country{}
}
\author{Jeremie S. Kim}
\affiliation{%
  \institution{ETH Z{\"u}rich}
  \country{}
}
\author{Onur Mutlu}
\affiliation{%
  \institution{ETH Z{\"u}rich}
  \country{}
}
\renewcommand{\shortauthors}{Orosa and Ya{\u{g}}l{\i}k{\c{c}}{\i}, et al.}


\sloppypar
\input{00_abstract}



\setlength{\footskip}{35pt} 

\fancyfoot[C]{\fontsize{11pt}{11pt}\selectfont\thepage} 
\renewcommand{\headrulewidth}{0pt}
\renewcommand{\footrulewidth}{0pt}

\fancypagestyle{plain}{
  \fancyhf{}
  \fancyhead[C]{}     
  \fancyfoot[L]{This is a notice}
  \renewcommand{\footrulewidth}{0pt} 
  \renewcommand{\headrulewidth}{0pt}
}

\maketitle

\pagestyle{fancy}

\AddToShipoutPictureBG*{%
  \AtPageLowerLeft{%
    \setlength\unitlength{1in}%
    \hspace*{\dimexpr0.5\paperwidth\relax}
    \makebox(0,1.05)[c]{\thepage}%
}}


\input{01_introduction}

\input{02_background}

\input{03_motivation}
\input{04_methodology}
\input{05_temperature}
\input{06_temporal}
\input{07_spatial}
\input{08_implications}
\input{09_related}
\input{10_conclusion}
\begin{acks} 
{We thank the anonymous reviewers of MICRO~2021 for {feedback}. We thank the SAFARI Research Group members for {valuable} feedback and the stimulating intellectual environment they provide. We acknowledge the generous gifts provided by our industrial partners: Google, Huawei, Intel, Microsoft, and VMware.}
\end{acks}


\balance
\bibliographystyle{ACM-Reference-Format} 
\bibliography{refs}

\onecolumn
\appendix

\setcounter{table}{3}
\input{11_appendix}



\end{document}

%% file: macros.tex
\definecolor{amber}{rgb}{1.0, 0.49, 0.0}
\definecolor{awesome}{rgb}{1.0, 0.13, 0.32}
\definecolor{dollarbill}{rgb}{0.52,0.73,0.4}
\definecolor{moegi}{rgb}{0.357, 0.537, 0.188}
\definecolor{burgundy}{rgb}{0.5, 0.0, 0.13}
\definecolor{ballblue}{rgb}{0.13, 0.67, 0.8}
\definecolor{ups-truck}{rgb}{0.53, 0.28, 0.21}
\definecolor{airforceblue}{rgb}{0.36, 0.54, 0.66}
\definecolor{cadmiumgreen}{rgb}{0.0, 0.42, 0.24}
\definecolor{darkcyan}{rgb}{0.0, 0.55, 0.55}
\definecolor{caribbeangreen}{rgb}{0.0, 0.8, 0.6}
\definecolor{flamingopink}{rgb}{0.99, 0.56, 0.67}
\definecolor{jazzberryjam}{rgb}{0.65, 0.04, 0.37}
\definecolor{mediumpersianblue}{rgb}{0.0, 0.4, 0.65}
\definecolor{coolblack}{rgb}{0.0, 0.18, 0.39}
\definecolor{bleudefrance}{rgb}{0.19, 0.55, 0.91}
\definecolor{ao}{rgb}{0.0, 0.0, 1.0}
\definecolor{babyblueeyes}{rgb}{0.63, 0.79, 0.95}
\definecolor{antiquefuchsia}{rgb}{0.57, 0.36, 0.51}

\newcommand{\squishlist}{
 \begin{list}{$\circ$}
  { \setlength{\itemsep}{0pt}
     \setlength{\parsep}{0pt}
     \setlength{\topsep}{0pt}
     \setlength{\partopsep}{0pt}
     \setlength{\leftmargin}{1em}
     \setlength{\labelwidth}{1em}
     \setlength{\labelsep}{0.5em} } }

\newcommand{\squishsublist}{
\begin{list}{$\rightarrow$}
 { \setlength{\itemsep}{0pt}
    \setlength{\parsep}{0pt}
    \setlength{\topsep}{-10em}
    \setlength{\partopsep}{-3pt}
    \setlength{\leftmargin}{1em}
    \setlength{\labelwidth}{1em}
    \setlength{\labelsep}{0.5em} } }

\newcommand{\squishend}{
  \end{list}  }

\newcounter{obs}
\newcommand\observation[1]{%
   \refstepcounter{obs}
   \noindent
   \colorbox{gray!20}{\textbf{\obslbl{} \theobs.}} \emph{#1}}
   
\newcounter{take}
\newcommand\take[1]{%
   \refstepcounter{take}
  \vspace{0.5em}
  \noindent
  \begin{tabular}{|p{0.95\linewidth}|}
       \hline
       \textbf{{Takeaway \thetake}.} \emph{{#1}}\\
       \hline 
  \end{tabular}
  \vspace{-0.5em}
  \addtocounter{table}{-1} 
}

\newcommand{\figlbl}[0]{Fig.}
\newcommand{\figref}[1]{\figlbl{}~\ref{#1}}

\newcommand{\figlbls}[0]{Figs.}
\newcommand{\figrefs}[1]{\figlbls{}~\ref{#1}}

\newcommand{\seclbl}[0]{§}
\newcommand{\secref}[1]{\seclbl{}\ref{#1}}

\newcommand{\obslbl}[0]{Obsv.}
\newcommand{\obsref}[1]{\obslbl{}~\ref{#1}}

\newcommand{\obslbls}[0]{Obsvs.}
\newcommand{\obsrefs}[1]{\obslbls{}~\ref{#1}}

\newcommand{\vshort}[0]{\vspace{-6pt}}

\newif\ifcameraready
\camerareadytrue
\camerareadyfalse

\newif\ifrevision
\revisionfalse

\newif\ifsubmission
\submissionfalse

\newif\ifonurready
\onurreadytrue

\ifcameraready

    \newcommand{\om}[1]{{#1}}

    \newcommand{\atbc}[1]{}
    \newcommand{\loiscomment}[1]{}
    \newcommand{\onurcomment}[1]{}
    \newcommand{\agycomment}[1]{}
    \newcommand{\hluocomment}[1]{}
    \newcommand{\jktodo}[1]{}
    \newcommand{\mptodo}[1]{}
    
    \newcommand{\agyl}[1]{}
    \newcommand{\oml}[1]{}
    \newcommand{\hluol}[1]{}
    \newcommand{\param}[1]{{#1}}
    \newcommand{\ignore}[1]{}

\else
\ifonurready 

    \newcommand{\om}[1]{\textcolor{jazzberryjam}{#1}}

    \newcommand{\atbc}[1]{\textcolor{blue}{\textbf{[@atb: #1]}}}
    \newcommand{\loiscomment}[1]{\noindent{\color{blue}{\bf \fbox{Lois: }}{\it#1}}}
    \newcommand{\onurcomment}[1]{\noindent{\color{blue}{\bf \fbox{ONUR: }}{\it#1}}}
    \newcommand{\agycomment}[1]{\textcolor{blue}{\textbf{[@gy:~{\it#1}]}}}

    \newcommand{\hluocomment}[1]{\textcolor{blue}{\textit{[@hluo: #1]}}}

    \newcommand{\agyl}[1]{\todo[size=\scriptsize,color=babyblueeyes]{\textbf{{@gy:}} #1}}
    \newcommand{\oml}[1]{\todo[size=\scriptsize,color=babyblueeyes]{\textbf{{onur:}} #1}}
    
    \newcommand{\hluol}[1]{\todo[size=\scriptsize,color=babyblueeyes]{\textbf{{@hluo:}} #1}}
    
    \newcommand{\jktodo}[1]{\textcolor{blue}{\textbf{[JKTODO:}#1\textbf{]}}}
    \newcommand{\mptodo}[1]{\textbf{\scriptsize \fcolorbox{black}{red}{\color{white}{TODO}}}\textcolor{red}{\emph{#1}}} 
    
    \newcommand{\param}[1]{{#1}}

\else
    \ifrevision

      \newcommand{\agyl}[1]{\todo[size=\small,color=babyblueeyes]{\textbf{{@gy:}} #1}}
      \newcommand{\hluol}[1]{\todo[size=\scriptsize,color=babyblueeyes]{\textbf{{@hluo:}} #1}}

      \newcommand{\param}[1]{{#1}}
      \newcommand{\om}[1]{\textcolor{coolblack}{#1}}

      \newcommand{\agycomment}[1]{}
      \newcommand{\loiscomment}[1]{}
      \newcommand{\onurcomment}[1]{}
      \newcommand{\hluocomment}[1]{}
      \newcommand{\jktodo}[1]{}
      \newcommand{\mptodo}[1]{}
      \newcommand{\atbc}[1]{}
      \newcommand{\ignore}[1]{}
      
    \else

      \ifsubmission   

        \newcommand{\param}[1]{#1}
        
        \newcommand{\om}[1]{#1}

        \newcommand{\agycomment}[1]{}
        \newcommand{\loiscomment}[1]{}
        \newcommand{\onurcomment}[1]{}
        \newcommand{\hluocomment}[1]{}
        \newcommand{\atbc}[1]{}
        \newcommand{\ignore}[1]{}
        \newcommand{\jktodo}[1]{}
        \newcommand{\mptodo}[1]{}
        
        \newcommand{\agyl}[1]{}

    \else           
        
        \newcommand{\jktodo}[1]{\textcolor{mediumpersianblue}{\textbf{[JKTODO:}#1\textbf{]}}}
        \newcommand{\mptodo}[1]{\textbf{\scriptsize \fcolorbox{black}{red}{\color{white}{TODO}}}\textcolor{red}{\emph{#1}}} 
        
        \newcommand{\param}[1]{\textcolor{ballblue}{#1}}
        \newcommand{\om}[1]{\textcolor{jazzberryjam}{#1}}

        \newcommand{\atbc}[1]{\textcolor{ups-truck}{\textbf{[@atb: #1]}}}

        \newcommand{\loiscomment}[1]{\noindent{\color{red}{\bf \fbox{Lois: }}{\it#1}}}
        \newcommand{\onurcomment}[1]{\noindent{\color{red}{\bf \fbox{ONUR: }}{\it#1}}}
        \newcommand{\agycomment}[1]{\textcolor{awesome}{\textbf{[@gy: {\it#1}]}}}
        \newcommand{\hluocomment}[1]{\textcolor{moegi}{\textit{[@hluo: #1]}}}
        \newcommand{\ignore}[1]{}
        \newcommand{\agyl}[1]{\todo[size=\scriptsize,color=babyblueeyes]{\textbf{{@gy:}} #1}}
        \newcommand{\oml}[1]{\todo[size=\scriptsize,color=babyblueeyes]{\textbf{{onur:}} #1}}
        \newcommand{\hluol}[1]{\todo[size=\scriptsize,color=babyblueeyes]{\textbf{{@hluo:}} #1}}
    \fi
  \fi
\fi
\fi

\newcommand{\tras}{t_{RAS}}
\newcommand{\trp}{t_{RP}}

\newacronym{hcfirst}{\textit{HC\textsubscript{first}}}{the minimum hammer count {value} at which the first bit error is observed}
\newacronym{ber}{$BER$}{the number of bit flips in a DRAM row, refer{red} to as bit error rate}
\newacronym{wcdp}{$WCDP$}{worst-case data pattern}

\newcommand{\numchips}{248}
\newacronym{taggon}{$t_{AggOn}$}{the time that an aggressor row stays active, {i.e., aggressor row's on-time}}
\newacronym{taggoff}{$t_{AggOff}$}{the time that {the bank} stays precharged, {i.e., aggressor row's off-time}}
\newacronym{tras}{$\tras{}$}{the minimum time that a row should stay active before a precharge command is issued {to the bank}}
\newacronym{trp}{$\trp{}$}{the minimum time a precharge command needs to complete before a row is activated {in the same bank}}
\newacronym{iqr}{$IQR$}{interquartile range}
\newacronym{cv}{$CV$}{the coefficient of variation}
\newacronym{hc}{$HC$}{hammer count}
\newacronym{rfm}{$RFM$}{refresh management}

%% file: 00_abstract.tex
\begin{abstract}
RowHammer is a circuit-level DRAM vulnerability where repeatedly accessing (i.e., hammering) a DRAM row can cause {bit flips} in physically nearby rows.
The RowHammer vulnerability worsens as DRAM cell size and cell-to-cell spacing shrink.
Recent studies demonstrate that modern DRAM chips, including chips previously marketed as RowHammer-safe, are even more vulnerable to RowHammer than older chips such that the required hammer count to cause a {bit flip} has reduced by more than 10X in the last decade.
Therefore, it is essential to develop a better understanding and in-depth insights into the RowHammer vulnerability of modern DRAM chips to more effectively secure current and future systems.

Our goal in this paper is to provide insights into fundamental properties of the RowHammer vulnerability that are not yet rigorously studied by prior works, but can potentially be $i$) exploited to develop more effective RowHammer attacks or $ii$) leveraged to design more effective and {efficient} defense mechanisms.
To this end, we present an experimental characterization {using} \param{\numchips{}}~DDR4 {and \param{24}~DDR3 {modern} DRAM chips} from four major DRAM manufacturers demonstrating how the RowHammer effects vary with three fundamental properties: 1)~{DRAM chip} temperature, 2)~{aggressor} row active time, and 3)~{victim} DRAM cell's physical location.
{Among our \param{16} new observations, we highlight} that a RowHammer {bit flip} 1)~is {very likely} to occur {in a bounded} range, {specific {to each DRAM cell}} {(e.g., 5.4\% of the vulnerable DRAM cells exhibit errors in the range \SI{70}{\celsius} to \SI{90}{\celsius})}, 2)~is more likely to occur if  the aggressor row {is} active for longer time {(e.g., RowHammer vulnerability {increases} by 36\% {if we} keep a DRAM row active for 15 column accesses)}, and 3)~is more likely to occur in {certain physical regions} of {the} DRAM module {under attack} {(e.g., 5\% of the rows are 2x more vulnerable than the remaining 95\% of the rows)}.
{Our study has important practical implications on future RowHammer attacks and defenses. We {describe} and analyze the implications of our new findings by proposing \param{three} future RowHammer attack and \param{{six}} future {RowHammer} defense improvements.}
\end{abstract}

%

%% file: 01_introduction.tex
\section{Introduction}
\label{sec:intro}
To maintain competitive DRAM prices, manufacturers focus on reducing the cost-per-bit of DRAM via DRAM circuit designs and manufacturing process technology that improve DRAM storage density, which in turn reduces DRAM cell size and cell-to-cell spacing. Unfortunately, these reductions have been shown to negatively impact DRAM reliability~{\cite{mutlu2013memory, meza2015revisiting}} and expose vulnerabilities {such as} RowHammer~\cite{kim2014flipping, kim2020revisiting, mutlu2019rowhammer}. RowHammer is an error mechanism that is caused by \emph{hammering}, or opening and closing (i.e., \emph{activating} and \emph{precharging}), a DRAM row {(i.e., \emph{aggressor row})} {many times}, which can cause {bit flips} in physically-nearby rows (i.e., \emph{victim rows}){~\cite{ kim2014flipping, mutlu2017rowhammer, yang2019trap, mutlu2019rowhammer,park2016statistical, park2016experiments,
walker2021ondramrowhammer, ryu2017overcoming, yang2016suppression, yang2017scanning, gautam2019row, jiang2021quantifying}}. RowHammer has gained attention in both academia and industry, and various attacks have {exploited} the RowHammer vulnerability to escalate {privilege}, leak private data, and manipulate critical application outputs~\cite{seaborn2015exploiting, van2016drammer, gruss2016rowhammer, razavi2016flip, pessl2016drama, xiao2016one, bosman2016dedup, bhattacharya2016curious, qiao2016new, jang2017sgx, aga2017good, mutlu2017rowhammer, tatar2018defeating, gruss2018another, lipp2018nethammer, van2018guardion, frigo2018grand, cojocar2019eccploit,  ji2019pinpoint, mutlu2019rowhammer, hong2019terminal, kwong2020rambleed, frigo2020trrespass, cojocar2020rowhammer, weissman2020jackhammer, zhang2020pthammer, rowhammergithub, yao2020deephammer, hassan2021utrr}.\footnote{A survey of RowHammer studies and attacks can be found in~\cite{mutlu2019rowhammer}.}
To make matters worse, recent experimental studies~\cite{mutlu2017rowhammer, mutlu2019rowhammer, frigo2020trrespass, cojocar2020rowhammer, kim2020revisiting, kim2014flipping} have found that the RowHammer vulnerability is {becoming more severe} in newer DRAM chip generations. {For example, as shown in~\cite{kim2020revisiting}}, chips manufactured in 2020 can experience RowHammer {bit flips} after an order of magnitude fewer row activations compared to the chips manufactured in 2014~\cite{kim2014flipping}. As the RowHammer vulnerability worsens, ensuring RowHammer-safe operation {becomes} more expensive in terms of performance overhead, energy consumption, and hardware complexity~\cite{kim2020revisiting, yaglikci2021blockhammer, park2020graphene}. Therefore, it is {critical} to understand RowHammer in greater detail with in-depth insights into how the RowHammer vulnerability varies under different conditions in order to develop more effective and {efficient} solutions for the security and reliability of current and future DRAM-based computing systems.

Our goal in this paper is to provide insights into fundamental properties of the RowHammer vulnerability that are not yet rigorously studied by prior works, but can potentially be $i$) exploited to develop more effective RowHammer attacks or $ii$) leveraged to design more effective and {efficient} defense mechanisms.
To this end, we provide a rigorous experimental characterization {of} \param{\numchips{}}~DDR4 {and \param{24}~DDR3 {modern} DRAM chips} from \param{four} major manufacturers
to understand how RowHammer vulnerability changes with three fundamental properties of a RowHammer attack: {1)~DRAM chip temperature, 2)~aggressor row active time, and 3)~victim DRAM cell's physical location.} This is the first paper that rigorously analyzes these \param{three} properties.

Based on our novel characterization results, we make {\param{16} new} observations {and share \param{6} key takeaway lessons from our observations. We leverage these observations to propose {\param{three} RowHammer attack and \param{six} RowHammer defense improvements}}.
From our {\param{16} new} observations, {we highlight} three observations that are especially important.
First, we find that each vulnerable DRAM cell can experience a RowHammer {bit flip} {only} within a bounded temperature range. This range can be as narrow as \SI{5}{\celsius} or as wide as \SI{40}{\celsius} (in our tested chips). 
{Second, when the aggressor row's active time is longer (e.g., by 5$\times$), 1)~more DRAM cells (6.9$\times$ on average) experience RowHammer bit flips at a given hammer count and 2)~a DRAM row experiences RowHammer bit flips at a smaller hammer count (by 36\% on average).}
Third, {a small fraction of} DRAM {rows} in a DRAM module {(5\%/1\%)} are \emph{significantly {(2{.0}$\times$/1.6$\times$)} more vulnerable} to RowHammer than {{the rest (95\%/99\%) of}} the module.

To study RowHammer effects at the circuit level, we disable RowHammer mitigation mechanisms {in the real DRAM chips we characterize}. For each experiment, we 1) perform a double-sided RowHammer attack~\cite{kim2014flipping, seaborn2015exploiting, kim2020revisiting}, in which both physically-adjacent aggressor rows of the victim row are repeatedly accessed (i.e., hammered), and 2) maintain a high-precision 
{(i.e., error of {at} most $\pm$\SI{0.1}{\celsius})}
temperature-controlled environment for DRAM. 
We conduct \param{{three}} main {analyses} in our characterization study. 

{First, we investigate the effects of temperature on {both 1)~\gls{ber} and~2)~\gls{hcfirst} in} a victim DRAM row under RowHammer attack.
{Our \gls{ber} analysis} demonstrates that a vulnerable DRAM cell experiences bit flips in a specific {and bounded} {range} of temperature, which can be {as narrow as} \SI{5}{\celsius}, or {as wide as} \SI{40}{\celsius}.
{Our \gls{ber} analysis also shows} that the effect of temperature on the \gls{ber} of a DRAM chip highly depends on the DRAM chip manufacturer. For example, DRAM chips of one manufacturer show increasing \gls{ber} with temperature, whereas DRAM chips from another manufacturer show decreasing \gls{ber} with temperature. 
{Our analysis {of} \gls{hcfirst}} demonstrates that the RowHammer vulnerability tends to worsen as temperature {increases}.}

{{Second}}, we test the sensitivity of RowHammer {bit flips} to the active time of an aggressor row. To do so, we change the time between an aggressor row activation to the {succeeding} precharge command from \param{\SI{34.5}{\nano\second} to \SI{154.5}{\nano\second} with \SI{30}{\nano\second} steps} while the total hammer count is fixed at {a given value. Using this methodology we analyze the variation in both \gls{hcfirst} and \gls{ber}.} We observe that as the time between the aggressor row activation and precharge command increases, DRAM cells become more vulnerable to RowHammer.

{Third}, we analyze how {RowHammer vulnerability varies} based on the {\emph{physical location}} of a DRAM cell. We observe that {\gls{hcfirst} significantly varies across rows such that {only a small fraction of DRAM rows (}{\param{5\%/1\%}) exhibit {significantly higher RowHammer vulnerability (}\param{2.0$\times$/1.6$\times$} lower \gls{hcfirst} values {on average across all four manufacturers}) than the {rest of the rows (\param{95\%/99\%}).}}}

{Based on our {new} observations, we {describe and} analyze \param{three {(six)}} {improvements to increase the effectiveness of existing} RowHammer attacks {(defense mechanisms).}}

We make the following contributions in this work:

\squishlist{}
    \item We present the first rigorous experimental study that examines temperature effects on RowHammer {bit flips} in modern DRAM chips. 
    {Our tests using} \param{\numchips{}}~DDR4 {and \param{24}~DDR3 {modern} DRAM chips} from four major {manufacturers} demonstrate that {a DRAM cell experiences bit flips in a specific and bounded range of temperature and the RowHammer vulnerability tends to worsen as temperature {increases}.} 
    
    \item {We} experimentally demonstrate, {for the first time,} how RowHammer {vulnerability} changes with the active time of the aggressor rows. {Our results show that as the} aggressor row's active time {increases} {(e.g., by 5$\times$), 1)~more DRAM cells (6.9$\times$ on average) experience RowHammer bit flips at a given hammer count and 2)~a DRAM row experiences RowHammer bit flips at a smaller hammer count (by 36\% on average).}
    
    \item {We} demonstrate that a DRAM cell's RowHammer vulnerability significantly depends on the cell's location. 
    We observe that {{only a small fraction of DRAM rows (\param{5\%/1\%}{)} exhibit significantly higher RowHammer vulnerability (\param{2.0$\times$/1.6$\times$} lower \gls{hcfirst} values) than the {rest (\param{95\%/99\%}) of the rows.}}}

    \item {Based on our {new} observations {on RowHammer's sensitivities to temperature, aggressor row's active time, and a victim DRAM cell's physical location in the DRAM chip}, we {describe and} analyze {three} future RowHammer attack {and six future RowHammer} defense {improvements}.} 
\squishend{}

%% file: 02_background.tex
\section{Background}
\label{sec:background}

We {provide a brief} background on DRAM organization, DRAM access timings, and RowHammer vulnerability. 
{For more detailed background on the{se}, we refer the reader to many prior works~{\cite{liu2012raidr, liu2013experimental, keeth2001dram, mutlu2007stall, moscibroda2007memory, mutlu2008parbs, kim2010atlas, subramanian2014bliss, salp, kim2014flipping, qureshi2015avatar, hassan2016chargecache, chang2016understanding, lee2017design,  chang2017understanding,  patel2017reaper,kim2018dram, kim2020revisiting, hassan2019crow, frigo2020trrespass, chang2014improving, chang2016low, vampire2018ghose, hassan2017softmc, khan2016parbor, khan2016case, khan2014efficacy, seshadri2015gather, seshadri2017ambit, kim2018solar, kim2019d, patel2019understanding, patel2020beer, lee2013tiered, lee2015decoupled, seshadri2013rowclone, luo2020clrdram, seshadri2019dram, wang2020figaro,orosa2021codic,wang2018reducing,ipek2008self,zhang2014half}}}. 

\subsection{DRAM Organization}
\label{sec:background_dramorg}
\begin{figure}[t] \centering
    \includegraphics[width=\linewidth]{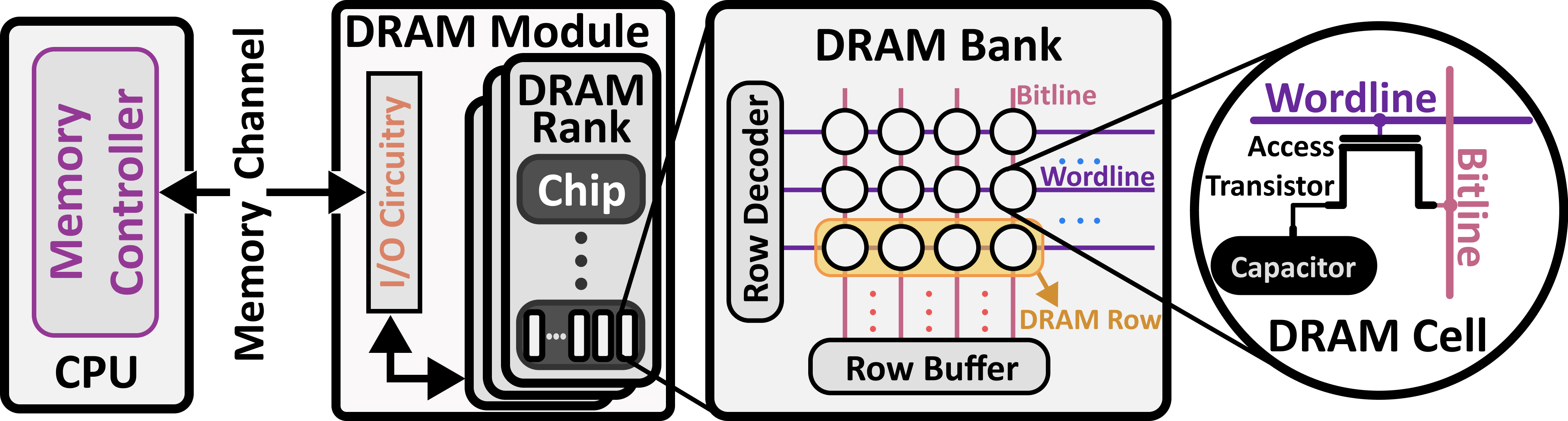}
    \caption{DRAM organization.}
    \label{fig:dram-organization}
    \vshort{}
\end{figure}

\figref{fig:dram-organization} depicts the {hierarchical} organization of DRAM-based main memory.
{T}he \emph{memory controller} in a system is {typically} connected to
multiple DRAM \emph{modules} {via} 
multiple DRAM \emph{channels}.
Each channel operates independently. 
A DRAM \emph{module} {has one or more ranks{,} {each of which} consist{ing} of} multiple DRAM \emph{chips} {that} operate in lock-step. 
The memory controller {can interface with multiple DRAM ranks by time-multiplexing the channel’s I/O bus between the ranks. Because the I/O bus is shared, the memory controller serializes accesses to different ranks in the same channel.}
A DRAM chip is organized into multiple DRAM \emph{banks}. DRAM banks in a DRAM chip share a common I/O circuitry.

DRAM \emph{cells} in a DRAM bank are laid out {in} a two-dimensional structure {of \emph{rows}} and \emph{columns}. Each DRAM cell on a DRAM row is connected to a {common \emph{wordline}} via \emph{access transistors}. {A \emph{bitline}} connects a column of DRAM cells to a DRAM sense amplifier to access and manipulate data. The two-dimensional array structure of DRAM cells is typically partitioned into multiple DRAM \emph{subarrays}~\cite{salp, chang2014improving, seshadri2013rowclone}. {Each subarray is connected to sense amplifiers that enable sensing of data{,} called \emph{local row buffers}.} 

\subsection{DRAM Access Timings}

The memory controller accesses DRAM {locations via} three {major} steps. First, the memory controller issues an $ACT$ command to activate a specific row within a bank, which prepares the row for a column access. Second, the memory controller issues a $RD$ or $WR$ command to read or write to a column in the row, respectively. Third, once all column operations to the active row are complete, the memory controller issues a {precharge (}$PRE${)} command, which closes the row and prepares the bank for a new activation. 

To {guarantee} correct DRAM operation, the memory controller must observe standardized timings between {consecutive} commands{, called} \emph{timing parameters}~\cite{salp, lee2013tiered, lee2015adaptive}{. Timing parameters} ensure that the internal DRAM circuitry has sufficient time to perform {the operations} required by the command. In this work, we deal with two key timings: 1) \gls{tras} and 2) \gls{trp}.
{\gls{tras}} ensures that the DRAM sense amplifiers have enough time following a row activation to correctly restore the charge in all cells {in} the open row before the row is closed. 
{\gls{trp}} ensures that all bitlines in the subarray are fully precharged to their idle {reference} voltage (typically $V_{DD}/2$) before the next row is activated. 

\subsection{The RowHammer Vulnerability}
\label{sec:background_rowhammer}
Modern DRAM {chips} suffer from {an} error {mechanism, called RowHammer~\cite{kim2014flipping, mutlu2019rowhammer, kim2020revisiting}} that happens when a DRAM row {(i.e., aggressor row)} {is repeatedly activated {enough times before its neighboring rows {(i.e., victim rows) get} refreshed}}~\cite{kim2014flipping, mutlu2017rowhammer, yang2019trap, mutlu2019rowhammer,park2016statistical, park2016experiments,
walker2021ondramrowhammer, ryu2017overcoming, yang2016suppression, yang2017scanning, gautam2019row, kim2020revisiting, jiang2021quantifying}. {Due to the aggressive reduction in {manufacturing} process technology node size, DRAM cells become smaller and closer to each other, exacerbating the RowHammer vulnerability}. Therefore, as DRAM manufacturers continue to increase DRAM storage density, {DRAM chips'} vulnerability to RowHammer increases~\cite{mutlu2017rowhammer, mutlu2019rowhammer, mutlu2018rowhammer, frigo2020trrespass, cojocar2020rowhammer, kim2020revisiting,kim2014flipping}.

The RowHammer vulnerability can be used to
{reliably induce bit flips in main memory using various system-level security attacks~\cite{pessl2016drama, bosman2016dedup, bhattacharya2016curious, qiao2016new, jang2017sgx, aga2017good, mutlu2017rowhammer, van2018guardion, frigo2018grand, mutlu2019rowhammer, cojocar2019eccploit, cojocar2020rowhammer, frigo2020trrespass, weissman2020jackhammer, zhang2020pthammer, rowhammergithub, tatar2018defeating, gruss2016rowhammer, ji2019pinpoint, razavi2016flip, seaborn2015exploiting, van2016drammer, xiao2016one, gruss2018another, lipp2018nethammer, kwong2020rambleed, yao2020deephammer, hong2019terminal, hassan2021utrr}. Prior work demonstrates that inducing bit flips via {a} RowHammer attack is practical} for privilege escalation~\cite{gruss2018another,gruss2016rowhammer,ji2019pinpoint,lipp2018nethammer,razavi2016flip,seaborn2015exploiting,van2016drammer,xiao2016one}, denial of service~\cite{gruss2018another,lipp2018nethammer}, leaking confidential data~\cite{kwong2020rambleed}, {{and manipulating a critical application's correctness~\cite{yao2020deephammer, hong2019terminal}. Thus,} it is necessary to {rigorously} {understand the RowHammer vulnerability of modern DRAM chips, project future attacks, and develop} effective RowHammer {defense} mechanisms in modern systems that use DRAM.} 
{Through characterization~\cite{kim2014flipping, park2016experiments, park2016statistical, kim2020revisiting} and modeling~\cite{redeker2002investigation, park2016statistical, ryu2017overcoming, yang2016suppression, yang2017scanning, yang2019trap, gautam2019row, jiang2021quantifying, walker2021ondramrowhammer},
{past research shows} that circuit-level capacitive coupling~\cite{redeker2002investigation, jiang2021quantifying} and trap-assisted leakage~\cite{yang2019trap} have a significant effect on RowHammer {bit flips}~\cite{walker2021ondramrowhammer}.}

{Based on the understanding provided by prior characterization and modeling research,} a large body of research proposes various RowHammer defenses~\cite{AppleRefInc, kim2014flipping, kim2014architectural, aichinger2015ddr, bains2015row, aweke2016anvil, bains2016distributed, bains2016row, gomez2016dummy, jedec2017ddr4, brasser2017can, son2017making, konoth2018zebram, seyedzadeh2018cbt, van2018guardion, hassan2019crow, lee2019twice, kang2020cattwo, park2020graphene, yaglikci2021blockhammer, yaglikci2021security, ryu2017overcoming, yang2016suppression, yang2017scanning, gautam2019row, devaux2021method, you2019mrloc}. DRAM manufacturers {1})~implement RowHammer prevention mechanisms, {generally} called Target Row Refresh (TRR)~\cite{jedec2017ddr4, jedec2015hbm, frigo2020trrespass}, which perform proprietary operations within DRAM to {prevent} RowHammer {bit flips} {(without success, as shown by~\cite{frigo2020trrespass, hassan2021utrr})} {and 2) enhance DRAM communication protocols with {a new feature called \gls{rfm}}~\cite{jedec2020ddr5, jedec2020lpddr5}. {\gls{rfm} requires the memory controller to count the number of activations {at} DRAM bank granularity and issue a command when the activation count reaches a threshold value. By doing so, it provides an on-DRAM-die RowHammer defense mechanism (e.g., Silver Bullet~\cite{devaux2021method, yaglikci2021security})} with the necessary time to refresh victim rows. Despite {efforts to contain and defend against RowHammer, the vulnerability} {still exists and} is expected to worsen in the future~\cite{mutlu2017rowhammer, mutlu2018rowhammer, mutlu2019rowhammer, cojocar2020rowhammer, frigo2020trrespass, kim2020revisiting, hassan2021utrr}, {as clearly demonstrated by recent work~\cite{kim2020revisiting, frigo2020trrespass, hassan2021utrr}}}.

%% file: 03_motivation.tex
\section{Motivation and Goal}
\label{sec:motivation}
Prior research experimentally demonstrates that RowHammer is {clearly a} worsening DRAM reliability and security problem~\cite{kim2014flipping, mutlu2017rowhammer, mutlu2018rowhammer, mutlu2019rowhammer, cojocar2020rowhammer, frigo2020trrespass, kim2020revisiting}.
Despite all efforts, {newer} DRAM chips are shown to be \emph{significantly} more vulnerable to RowHammer than older generation chips~\cite{kim2020revisiting}. Even DRAM chips that have been marketed as RowHammer-free in 2020 experience RowHammer {bit flips} at \emph{significantly} lower hammer counts (e.g., \param{9.6K} for LPDDR4 chips when {TRR} protection is disabled~\cite{kim2020revisiting} and \param{{2}5K} {for DDR4 chips} when {TRR} protection is enabled~\cite{frigo2020trrespass}) compared to the {DDR3} DRAM chips manufactured in 2014 (e.g., \param{139K}~\cite{kim2014flipping}). Many prior works~\cite{aichinger2015ddr, AppleRefInc, aweke2016anvil, kim2014flipping, kim2014architectural,son2017making, lee2019twice, you2019mrloc, seyedzadeh2018cbt, van2018guardion, konoth2018zebram, park2020graphene, yaglikci2021blockhammer, kang2020cattwo, bains2015row, bains2016distributed, bains2016row, brasser2017can, gomez2016dummy, jedec2017ddr4,hassan2019crow, devaux2021method, ryu2017overcoming, yang2016suppression, yang2017scanning, gautam2019row, yaglikci2021security} have proposed RowHammer defense mechanisms to provide RowHammer-safe operation with either probabilistic or deterministic security guarantees. Unfortunately, {recent} works~\cite{kim2020revisiting, park2020graphene, yaglikci2021blockhammer} have demonstrated that many of these defense mechanisms will 
{incur} \emph{significant} performance, energy consumption, and/or hardware complexity overheads such that they become prohibitively expensive when deployed in future DRAM chips~\cite{kim2020revisiting}.

To enable RowHammer-safe operation in future DRAM-based computing systems in an effective {and} efficient way, it is {critical} to {rigorously} gain {detailed} insights into the RowHammer vulnerability {and its sensitivities to} varying attack properties. Unfortunately, despite the existing research efforts expended towards understanding RowHammer{~\cite{redeker2002investigation, ryu2017overcoming, yang2016suppression, yang2017scanning, yang2019trap, gautam2019row, jiang2021quantifying, walker2021ondramrowhammer, park2014active, park2016statistical, park2016experiments,kim2014flipping, kim2020revisiting}}, scientific literature lacks rigorous experimental observations on how {the} RowHammer vulnerability varies with three fundamental properties:
1)~DRAM chip {temperature}, 2)~{aggressor row active time}, 
and 3)~{victim DRAM cell's physical location}. This {lack of understanding} raises very practical and important concerns {{as to} how the {effects of these} three fundamental properties can be {exploited} to improve both RowHammer attacks and defense mechanisms.}

{\textbf{{O}ur goal}} in this paper is to {rigorously} evaluate and understand how the RowHammer vulnerability of {a} real DRAM chip at the circuit level changes with {1)~}temperature, {2)~aggressor row active time, and 3)~victim} DRAM cell{'s physical} location {in the DRAM chip}. Doing so provides us {with} a deeper understanding of RowHammer to enable future research {on improving the effectiveness of existing RowHammer attacks and defense mechanisms. We hope that these analyses will pave the way for {building} RowHammer-safe systems {that use increasingly more vulnerable} DRAM chips.}
{To achieve this goal, we rigorously characterize how the RowHammer vulnerability of \param{\numchips{}}~DDR4 {and \param{24}~DDR3 {modern} DRAM chips} from \param{four} major DRAM {manufacturers} vary with these three properties.}

%% file: 04_methodology.tex
\section{Methodology}
\label{sec:methodology}

We describe our methodology and infrastructure for characterizing the RowHammer vulnerability in real DRAM modules. 

\subsection{Testing Infrastructure}
\label{sec:testing_infrastructure}

We experimentally study {\param{\numchips{}}} DDR4 and \param{24} DDR3 DRAM chips across a {wide} range of {testing} conditions.
{We use two different testing infrastructures: 1)~SoftMC~\cite{hassan2017softmc, softmcgithub}, capable of precisely controlling temperature and command timings of DDR3 DRAM modules and 2)~a modified version of this infrastructure that supports DDR4 chips,}
also used in~\cite{kim2020revisiting, frigo2020trrespass, olgun2021quac, hassan2021utrr}.

\begin{figure}[htbp] \centering
    \includegraphics[width=\linewidth,trim={0 1.5cm 0 0},clip]{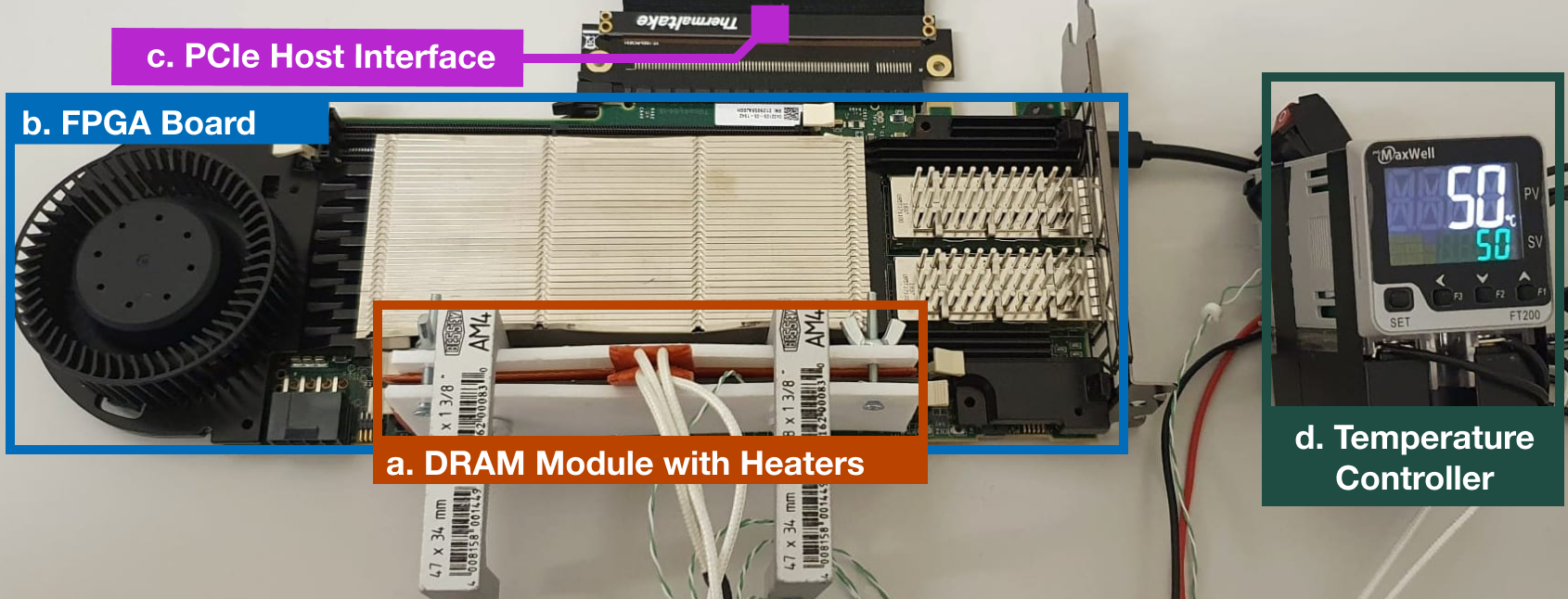}
    \vshort{}
    \vshort{}
    \caption{SoftMC Infrastructure: (a)~DRAM module under test clamped with heater pads, (b)~Xilinx Alveo U200 FPGA board~\cite{alveo}, programmed with a DDR4 version of SoftMC~\cite{hassan2017softmc}, (c)~PCIe connection to the host machine, and (d)~temperature controller.
    }
    \label{fig:infrastructure}
\end{figure}

\textbf{{SoftMC}.}
\figref{fig:infrastructure} shows one of our SoftMC setups for testing DDR4 modules (\figref{fig:infrastructure}a). 
We use {two} types of {Xilinx} FPGA boards: {1)}~Alveo U200~\cite{alveo} (\figref{fig:infrastructure}b) to test DDR4 DIMMs~\cite{jedec2017ddr4,micron2014ddr4}, and {2)}~ML605~\cite{ml605} to test DDR3 SODIMMs.
{This infrastructure enables precise control over both DDR4 and DDR3 timings at the granularity of \SI{1.25}{\nano\second} and \SI{2.50}{\nano\second}, respectively.}
We use a host machine, connected to our FPGA boards through a PCIe port~\cite{pcie} (\figref{fig:infrastructure}c) to
{1})~perform the RowHammer tests that we describe in \secref{sec:testing_methodology}
{and 2})~monitor and adjust the temperature of DRAM chips in cooperation with the temperature controller (\figref{fig:infrastructure}d).

\textbf{{Temperature Controller.}}
To regulate the temperature in DRAM modules, we use silicone rubber heaters pressed to both sides of the {DRAM} module (\figref{fig:infrastructure}a). 
We use a thermocouple, placed on the DRAM  chip to measure the chip's temperature {(similar to JEDEC standards~\cite{jedec1995thermal})}. A Maxwell FT200 temperature controller~\cite{maxwellFT200} (\figref{fig:infrastructure}d) 1)~monitors a DRAM chip's temperature using a thermocouple, and 2)~keeps the temperature stable by heating the chip with heater pads. The temperature controller 1)~communicates with our host machine 
via an RS485 channel~\cite{rs485} to get a reference temperature and to report the instant temperature, and 2) controls the heater pads using~a closed-loop PID controller.
In our tests using this infrastructure, we measure temperature with an error of {at} most $\pm$\SI{0.1}{\celsius}. {{We believe {that} our {temperature measurements from the DRAM package's surface accurately represent the DRAM die's real temperature} because the temperature of the DRAM package and the DRAM internal components are {strongly} correlated~\cite{micron2002temperature}.}}

\subsection{Testing Methodology}
\label{sec:testing_methodology}

\textbf{Disabling Sources of Interference.} 
Our goal is to directly observe the circuit-level bit flips such that we can make conclusions about DRAM’s vulnerability to RowHammer at the circuit technology level rather than {at} the system level. {To this end, we minimize all possible sources of interference with the following steps. 
First,} we disable all DRAM self-regulation events {(e.g., DRAM Refresh}~\cite{jedec2017ddr4,hassan2017softmc,ultrascale}) except {calibration related events (e.g., ZQ calibration} for signal integrity~\cite{jedec2017ddr4,hassan2017softmc}). {Second, we ensure that all RowHammer tests are conducted within a relatively short period of time such that we do not observe retention errors~\cite{liu2013experimental,khan2014efficacy,meza2015revisiting,patel2017reaper,qureshi2015avatar}}.  
{Third, we use the SoftMC memory controller~\cite{hassan2017softmc, softmcgithub} so that we can 1)~issue DRAM commands with precise control (i.e., our commands are not impeded by system-issued accesses), and 2)~study the RowHammer vulnerability on DRAM chips without interference from existing system-level RowHammer protection mechanisms (e.g.,~\cite{AppleRefInc, bains2015row, bains2016distributed, bains2016row}).
}

{Fourth, we test DRAM modules that do {\emph{not}} implement error correction codes (ECC){~\cite{hamming1950error,bose1960class,hocquenghem1959codes,reed1960polynomial,cojocar2019eccploit,kim2016all}}. {Doing so ensures} that {neither} on-die~\cite{patel2020beer,patel2019understanding,ECCMicron,nair2016xed,patel2021harp} {nor} rank-level~\cite{cojocar2019eccploit,kim2016all} ECC  {can} alter {the RowHammer bi{t f}lips we observe and analyze}.}
{Fifth}, we prevent known on-DRAM-die RowHammer defenses (i.e., TRR~\cite{jedec2020ddr5, jedec2020lpddr5,lee2014green,micron2016trr}) from working by not issuing refresh commands throughout our tests~\cite{frigo2020trrespass, kim2020revisiting}.

\textbf{{RowHammer}.} 
All our tests use double-sided {RowHammer}~\cite{kim2014flipping, kim2020revisiting, seaborn2015exploiting}, which activates, in an alternating manner, each of the two rows (i.e., aggressor rows) that are physically-adjacent to {a} victim row. {We call this victim row a double-sided victim row.} {We define single-sided victim rows as the rows that are hammered in a single-sided manner by the two aggressor rows (i.e., rows with +2 or -2 distance from victim row).} We define one hammer as a pair of activations to the two aggressor rows.
We perform double-sided {hammering} with the maximum activation rate possible within DDR3/DDR4 command timing {specifications}~\cite{jedec2017ddr4,jedec2012ddr3}. Prior works report that this is the most effective access {pattern} for RowHammer attacks on DRAM chips when {RowHammer mitigation mechanisms} are disabled~\cite{kim2014flipping, kim2020revisiting, frigo2020trrespass, cojocar2020rowhammer, seaborn2015exploiting}{.}{\footnote{Our analysis of aggressor row active time uses a different access sequence that introduces additional delays between row activations. {{S}ee \secref{sec:aggressor_active_time} for details.}}}
{W}e use 150K hammers (i.e., 300K activations) 
in {our} \gls{ber} experiments.\footnote{We find that 150K hammers is low enough to be used in a system-level RowHammer attack in a real system~\cite{frigo2020trrespass}, and it is high enough to provide a large number of bit flips in all DRAM modules we tested{.}} {We} use up to 512K hammers (i.e., the maximum number of hammers so that our hammer tests run for less than 64ms) in {our} \gls{hcfirst} experiments. Due to time limitations, {we repeat each test five times,} and we study the effects of the RowHammer attack on the 1) first \param{8K} rows, 2) last \param{8K} rows, and 3) middle \param{8K} rows of {a bank} in each DRAM chip (similar to ~\cite{kim2014flipping}).

\textbf{{Logical-to-Physical} Row Mapping.} 
DRAM manufacturers {use {DRAM-internal} mapping} {schemes} to internally translate memory-controller-visible row addresses to physical row {addresses~\cite{kim2014flipping, smith1981laser, horiguchi1997redundancy, keeth2001dram, itoh2013vlsi, liu2013experimental,seshadri2015gather, khan2016parbor, khan2017detecting, lee2017design, tatar2018defeating, barenghi2018software, cojocar2020rowhammer,  patel2020beer, yaglikci2021blockhammer}{,} which can vary across different DRAM modules}.
We {reverse-engineer} this mapping,
so {that} we can identify and hammer aggressor rows that are physically adjacent to a victim row. We reconstruct the mapping by 1) performing single-sided RowHammer attack on each DRAM row, 2) inferring that the two victim rows with the most RowHammer bit flips are physically adjacent to the aggressor row, and 3) deducing the address mapping after analyzing the aggressor-victim row relationships across all studied DRAM rows. 

\textbf{Data Pattern.} We conduct our experiments on a DRAM module by using the module's \gls{wcdp}.
We identify the \gls{wcdp} {for each module} as the pattern that results in the largest number of bit flips among \param{seven} different data patterns used in prior works on DRAM characterization~\cite{khan2014efficacy,liu2013experimental,patel2017reaper, kim2020revisiting, chang2016understanding, chang2017understanding, lee2017design,khan2016parbor,khan2016case,khan2017detecting,lee2015adaptive}{, presented in Table~\ref{tab:data_patterns}:} 
{colstripe, checkered, rowstripe{, and} random {(we also test the {complements} of the first three)}.} 
For each RowHammer test, we write the corresponding data pattern
to the victim row ($V$ in Table~\ref{tab:data_patterns}), and to the 8 previous ($V - [1...8]$) and next ($V + [1...8]$) physically-adjacent rows.

\begin{table}[h]
    \centering
    \vspace{-2pt}
    \footnotesize
    \caption{Data patterns used in our RowHammer {analyses.} }
    \vspace{-10pt}
    \begin{tabular}{l|cccc}
    \toprule
           \bf{Row {Address}}
           &  \bf{Colstripe{$^\dagger{}$}} & \bf{Checkered{$^\dagger{}$}} & \bf{Rowstripe{$^\dagger{}$}} & \bf{Random}\\
        \midrule
        $V^* \pm [0,2,4,6,8]$ &\verb$0x55$&\verb$0x55$ &\verb$0x00$&\verb$random$\\
        $V^* \pm [1,3,5,7]$ &\verb$0x55$&\verb$0xaa$&\verb$0xff$&\verb$random$\\
        \bottomrule
    \end{tabular}
    \begin{flushleft}
       $\quad^*V$ is the physical address of the victim row \\
       $\quad^\dagger{}${We also test the {complements} of these} patterns
    \end{flushleft}  
    \label{tab:data_patterns}
    \vshort{}
\end{table}

\glsresetall 

\textbf{Metrics.} We measure two metrics in our tests: 1)~\gls{hcfirst} and 2)~\gls{ber}. {A {\emph{lower}} \gls{hcfirst} {or \emph{higher} \gls{ber}} value indicates {higher} RowHammer vulnerability{.}}
{To {quickly} identify \gls{hcfirst}, we perform a binary search {where} we use an initial hammer count of 256k. We repeatedly increase (decrease) the hammer count by $\Delta$ if we observe {(do not observe)} bit flips in the victim row. The initial value is $\Delta = 128k$, and we halve it for each test until it reaches $\Delta = 512$ (i.e., we identify \gls{hcfirst} with an accuracy of 512 row activations){.}}

\textbf{Temperature {Range}.} To study the effects of temperature, we test DRAM chips across a wide range of temperatures, from \SI{50}{\celsius} to \SI{90}{\celsius}, with a step size of \SI{5}{\celsius}. 

\subsection{Characterized DRAM Chips}
\label{sec:characterized_dram_chips}
\label{sec:characterized_region}

Table~\ref{table:dram_chips} summarizes the \param{\numchips{}} DDR4 and \param{24} DDR3 DRAM chips we test from four major manufacturers. We use a {diverse set of} modules {with} different {chip} densities, die revisions and chip {organizations}. {We} share {analyses} of additional modules separately in~\cite{deepergithub}.

\begin{table}[h]
    \caption{{Summary of {DDR4 (DDR3)} {DRAM} chips tested.}}
    \vshort{}
    \centering
    \footnotesize{}
    \setlength\tabcolsep{3pt} 
    \begin{tabular}{cccrlcl}
        \toprule
            {{\bf Mfr.}} & \parbox{1.0cm}{\centering\bf{{DDR4}}\\\#DIMMs} & \parbox{1.2cm}{\centering\bf{{DDR3}}\\\#SODIMMs} & {{\bf  \#Chips}}  & {{\bf Density}} & {{\bf Die}}& {{\bf Org.}}  \\
        \midrule
        {Mfr. A}&9 &  1&144  (8)&{8Gb} (4Gb)&{B} (P)&{x4} (x8)\\
        {Mfr. B}&4 &  1& 32  (8)&{4Gb} (4Gb)&{F} (Q)&{x8} (x8)\\
        {Mfr. C}&5 &  1& 40  (8)&{4Gb} (4Gb)&{B} (B)&{x8} (x8)\\
        {Mfr. D}&4  & --& 32 (--)&{8Gb} (--)&{C} (--)&{x8} (--)      \\
        \bottomrule
    \end{tabular}
    \label{table:dram_chips}
\end{table}

%% file: 05_temperature.tex
\section{Temperature Analysis}
\label{sec:temperature}

\noindent
{We} {1)}~provide the first rigorous experimental characterization {of the effects of temperature} on the RowHammer vulnerability using real DRAM chips and {2)}~present new observations and insights based on our results.

\subsection{{Impact} of Temperature {on} DRAM Cells}

We analyze the relation between {temperature and the} RowHammer vulnerability of a DRAM cell {using the methodology described in Section~\ref{sec:testing_methodology}}. To {do so, {we first}} cluster vulnerable {DRAM} cells by their {\emph{vulnerable temperature range} (i.e., the {minimum and maximum temperatures within}} which a cell {experiences at least one} {RowHammer} bit flip across all {experiments}). 
{{Second,} we analyze {\emph{how}} the RowHammer bit flips of DRAM cells manifest within their vulnerable temperature range.}
Table~\ref{tab:temperature_gaps} shows the percentage of vulnerable cells that flip in \emph{all} temperature {points} of {their} vulnerable temperature {ranges}.

\begin{table}[h]
    \centering
    \footnotesize
    \caption{{Percentage of vulnerable DRAM cells that flip in all temperature points within the vulnerable temperature range of the cell.}}
    \vshort{}
    \begin{tabular}{rrrr}
        \toprule
        {\bf{Mfr. A}} & {\bf{Mfr. B}} & {\bf{Mfr. C}} & {\bf{Mfr. D}}\\
        \midrule
        {99.1\%} & {98.9\%} & {98.0\%} & {99.2\%} \\ 
        \bottomrule
    \end{tabular}
    \label{tab:temperature_gaps}
\end{table}

\observation{A DRAM cell is, with a very high probability, vulnerable to RowHammer in a continuous temperature range specific to the cell.\label{temp:bounded}}

For example, only \param{0.9\%} of the vulnerable DRAM cells in Mfr. A  do {\emph{not}} exhibit bit flips in at least one temperature point within their vulnerable temperature range. Hence, our experiments demonstrate that a cell {exhibits} bit flips  {with {very} high probability} in a {continuous} temperature range {that is specific to the cell}.

\begin{figure}[h] \centering
    \includegraphics[width=0.99\linewidth]{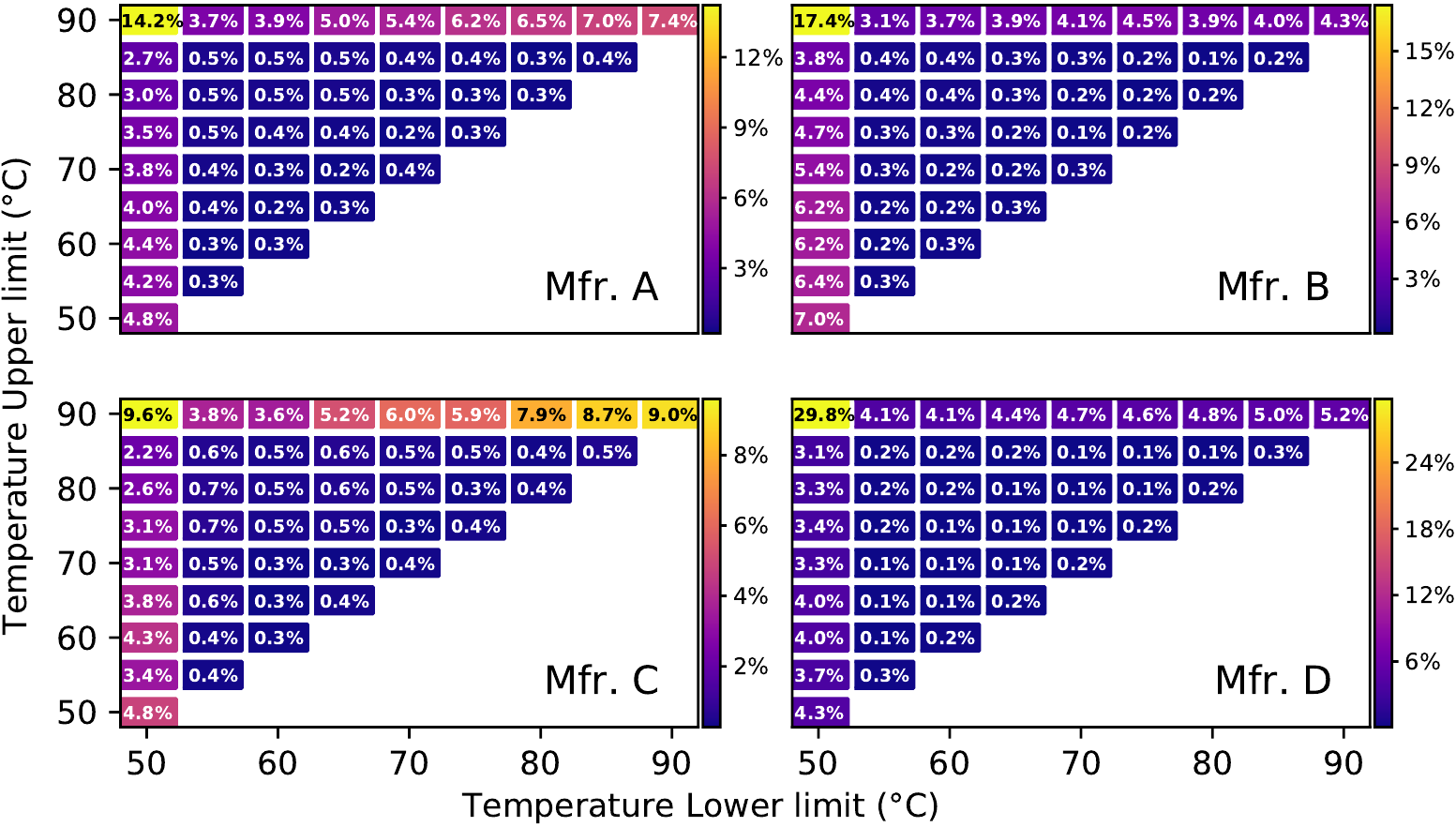}
    \vshort{}
    \caption{Population of {vulnerable DRAM cells}, clustered by vulnerable temperature range.}
    \label{fig:tempIntervals}
    \vshort{}
\end{figure}

{To analyze the diversity of vulnerable temperature ranges across DRAM cells, we cluster all vulnerable DRAM cells according to their vulnerable temperature {ranges}. \figref{fig:tempIntervals} shows each cluster's {size} as a percentage of the full population of vulnerable cells.}
{The x-axis (y-axis) indicates the lower (upper) bound of the vulnerable temperature range.}
{Because we do not test temperatures higher (lower) than \SI{90}{\celsius} (\SI{50}{\celsius}), the vulnerable temperature ranges with an upper (lower) limit of \SI{90}{\celsius} (\SI{50}{\celsius}) include cells that {also} flip  at higher (lower) temperatures. For example, 5.4\% of the vulnerable DRAM cells in Mfr. A fall into the range \SIrange{70}{90}{\celsius}, which includes cells with \emph{actual} vulnerable temperature ranges of} \SIrange{70}{95}{\celsius}, \SIrange{70}{100}{\celsius}, etc.

\observation{A significant fraction of vulnerable DRAM cells exhibit bit flips at all tested  temperatures.\label{temp:wide}}

We observe that {between \param{9.6\%} and \param{29.8\%} of the cells (x-axis}=\SI{50}{\celsius}, {y-axis}=\SI{90}{\celsius} in  \figref{fig:tempIntervals}) are vulnerable to RowHammer {across {\emph{all}} tested} temperatures {(\SIrange{50}{90}{\celsius})} for {the four} DRAM manufacturers. We also verify {(not shown)} that {\obsref{temp:wide}} holds for the three SODIMM DDR3 modules described in Table~\ref{table:dram_chips}.

\observation{A small fraction of all vulnerable DRAM cells are vulnerable to RowHammer only in a very narrow temperature range.\label{temp:narrow}}

For example, {\param{0.4\%} of all vulnerable DRAM cells {of {Mfr. A}},} are {only vulnerable to RowHammer at \param{\SI{70}{\celsius}} (i.e., a single tested temperature value).} {Note} that inducing even a single bit flip can be critical for system security, as shown by prior works~\cite{xiao2016one, frigo2018grand, gruss2018another, razavi2016flip}. Our experimental results show that 2.3\%, 1.8\%, 2.4\%, and 1.6\%  of all tested DRAM cells  for {Mrfs.} A, B, C, and D, respectively, experience a RowHammer bit flip within a temperature range as narrow as \SI{5}{\celsius}. {We conclude that {some} DRAM cells {experience} RowHammer bit flips at localized {and narrow} temperature {ranges}.}

{We exploit {\obsrefs{temp:bounded}--\ref{temp:narrow} in \secref{sec:implications}}.}

\take{{To ensure that a DRAM cell is not vulnerable to RowHammer, we must characterize the cell at all operating temperatures.}}

\subsection{Impact of Temperature on DRAM Rows}
\label{sec:temperature_rows}
We  analyze  the  relation  between  a DRAM row's RowHammer vulnerability and temperature in terms of both \gls{ber} and \gls{hcfirst}.

\begin{figure}[h] \centering
    \includegraphics[width=0.99\linewidth]{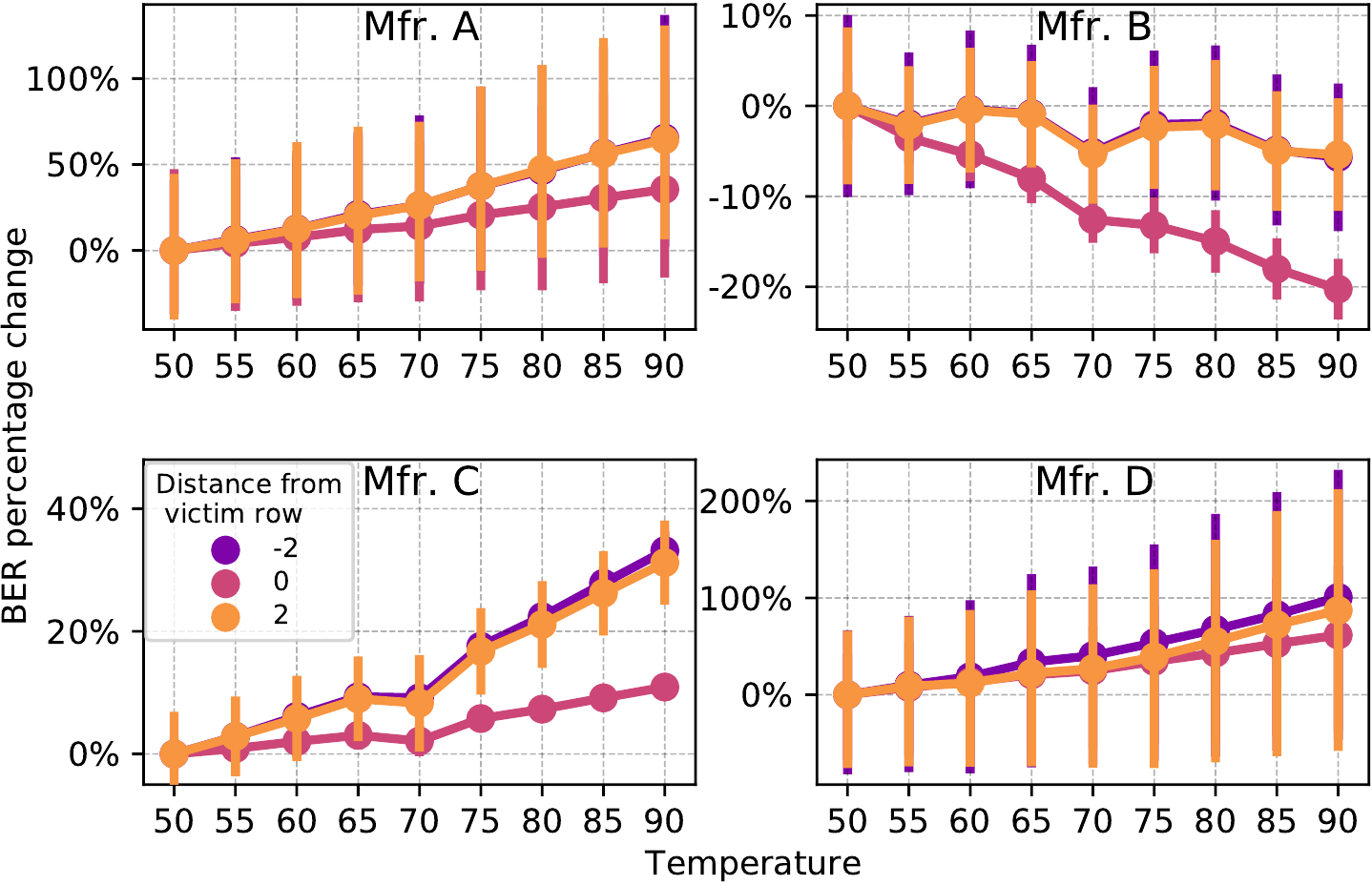}
    \vshort{}
    \caption{{{Percentage} change {in \gls{ber} (RowHammer bit flips)} with increasing temperature, compared to \gls{ber} at $50^\circ$C.}}
    \label{fig:ber_temp}
\end{figure}

\textbf{\emph{BER Analysis}.}
\figref{fig:ber_temp} shows {how the} \gls{ber} {changes} {as temperature increases}, {compared to the mean {\gls{ber}} value across all the samples at \SI{50}{\celsius}}, for four DRAM manufacturers. In each {plot}, we {use a point and error bar}\footnote{{Each point and error bar represent the mean and the 95\% confidence interval across the samples, respectively.}} to show the \gls{ber} {change} for the victim row (i.e., distance from the victim row = 0), and the \gls{ber} {change} for {the two single-sided victim rows (i.e., distance $\pm$2 from the victim row), across all rows we test}. 

\observation{{A DRAM row's} \gls{ber} can either increase or decrease with temperature depending on the DRAM manufacturer.\label{temp:ber_vs_temp}} 

{We observe that {the average \gls{ber} of} all three victim rows (one {double-sided} victim row and two single-sided victim rows),}
 from {Mfrs.} A, C, and D {increases with temperature}, whereas {the \gls{ber} of} rows from Mfr. B {decreases} as temperature increases. 
We hypothesize that the difference between these trends is caused by a combination of DRAM circuit design and manufacturing process technology differences {(see \secref{sec:temperature_circuit})}.

\begin{figure}[h] \centering
    \includegraphics[width=0.99\linewidth]{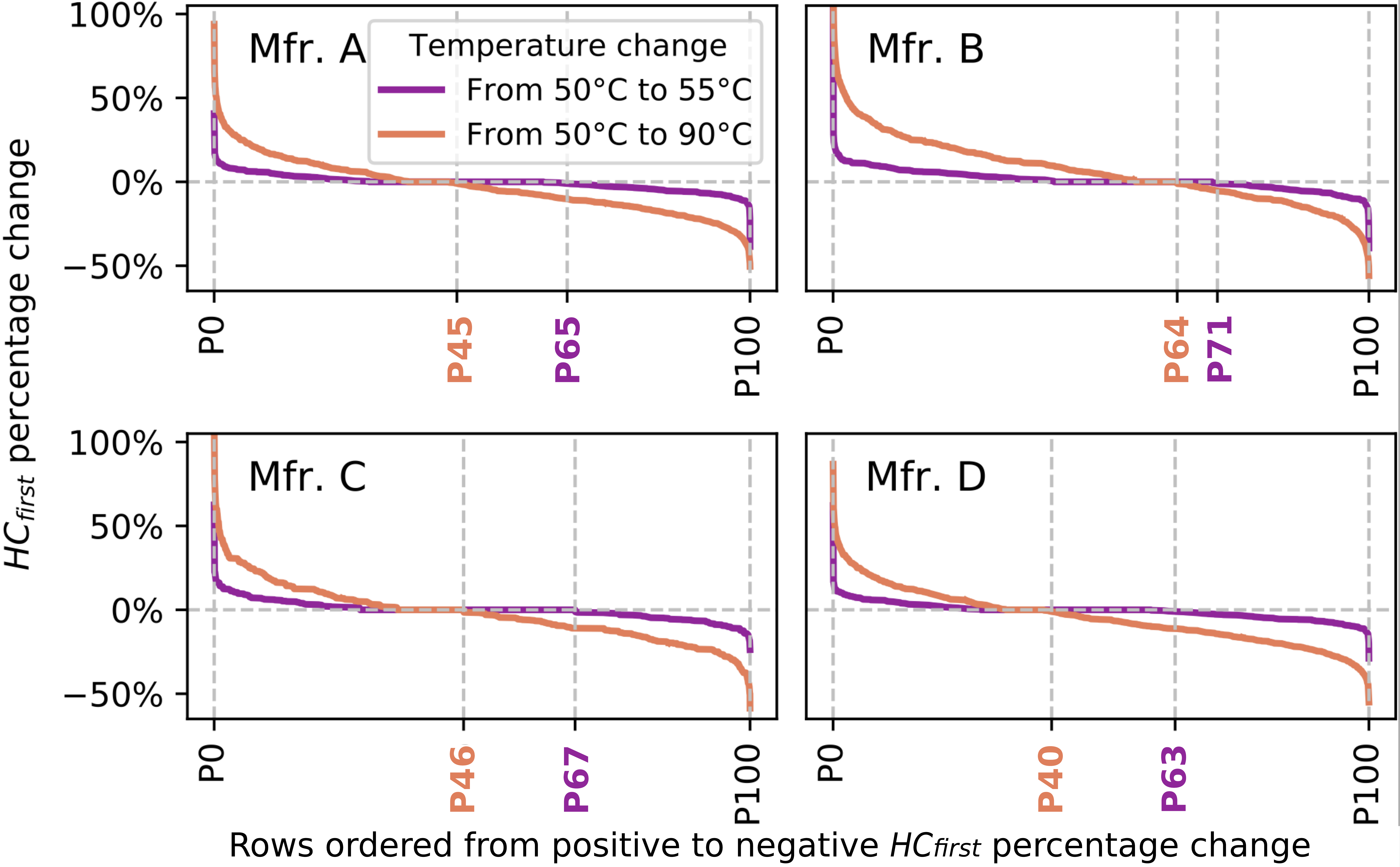}
        \vshort{}
    \caption{Distribution of {the change in} \gls{hcfirst} across vulnerable DRAM rows {as} temperature increases.}
    \vshort{}
    \label{fig:hcfirst_variation}
\end{figure}

\textbf{\gls{hcfirst} Analysis.}
\figref{fig:hcfirst_variation} shows the distribution of {the change in} \gls{hcfirst} {(in percentage)} when temperature {increases} from \param{\SI{50}{\celsius}} to \param{\SI{55}{\celsius}}, and from \param{\SI{50}{\celsius}} to \param{\SI{90}{\celsius}}, for the vulnerable rows of {the} \param{four} manufacturers. 
The x-axis {represents the percentage of all} vulnerable rows, sorted from {the} {most} positive \gls{hcfirst} {change} to {the} {most} negative \gls{hcfirst} {change}.
For each curve, we mark the {x-axis} {point at which the curve crosses the y=0\% line. This represents} the percentile of rows {whose} \gls{hcfirst} {increases with temperature;} 
e.g., for Mfr. A, when temperature {increases} from \SI{50}{\celsius} to {\SI{90}{\celsius}}, {only} 45\% (P45) of the tested rows have a higher \gls{hcfirst} {(indicating reduced vulnerability for that fraction of rows); i.e., most rows {from Mfr. A} are more vulnerable {at \SI{90}{\celsius} than at \SI{50}{\celsius}}}. For clarity, we only show two temperature changes (i.e., from \SI{50}{\celsius} to \SI{55}{\celsius} and from \SI{50}{\celsius} to \SI{90}{\celsius}), but our observations are consistent across all intermediate temperature changes we tested (i.e., from \SI{50}{\celsius} to 50+$\Delta ^\circ$C, for all $\Delta$'s that are {multiples} of \SI{5}{\celsius}). 

\observation{{DRAM rows can show either higher or lower \gls{hcfirst} when temperature increases.}\label{temp:hcfirst_vs_cells}}

{We observe that, for all four manufacturers, a significant {fraction} of rows can show either higher or lower \gls{hcfirst} when temperature increases. For example, when the temperature changes from  \SI{50}{\celsius} to \SI{55}{\celsius} in Mfr.~A, 65\% of the rows show higher \gls{hcfirst}, {while} 35\% of the rows show lower \gls{hcfirst}. We conclude that \gls{hcfirst} changes differently depending on the DRAM row.}

\observation{\gls{hcfirst} tends to generally decrease as temperature {change increases}.\label{temp:hcfirst_vs_temperature}}

We observe that, for all four manufacturers,
{fewer rows have a {higher} \gls{hcfirst} when the temperature delta is larger;} {i.e., the point at which each curve crosses the y=0\% {point} shifts left when the temperature change increases.}
For example, {for Mfr. D,} the fraction of {vulnerable} cells with {a higher} \gls{hcfirst} is much {larger} when temperature increases from \SI{50}{\celsius} to \SI{55}{\celsius} (63\% of cells) than when the temperature increases from \SI{50}{\celsius} to \SI{90}{\celsius} (40\% of  cells). {We conclude that the dominant trend is for a row's \gls{hcfirst} to decrease when {the temperature delta is larger}.} 

\observation{The change in \gls{hcfirst} tends to be larger as the temperature change is larger.\label{temp:hcfirst_vs_temperature_larger}}

{The \gls{hcfirst} distribution curve exhibits higher absolute magnitudes when temperature changes from \SI{50}{\celsius} to \SI{90}{\celsius}, compared to when temperature changes from \SI{50}{\celsius} to \SI{55}{\celsius} {(i.e., the curve {generally} rotates right {and has much higher peaks at its edges} when the temperature change increases, {i.e., going from orange to purple in the figure})}. {We quantify this observation by calculating the cumulative magnitude change (i.e., the sum of the absolute values of {the} \gls{hcfirst} change from all rows). Our results show that the cumulative magnitude change {(not shown in the figure)} is  4.2$\times$, 3.9$\times$, 3.8$\times$ and 4.3$\times$ larger in Mfrs. A, B, C, and D, respectively, when the temperature changes from \SI{50}{\celsius} to \SI{90}{\celsius}, compared to
\SI{50}{\celsius} to \SI{55}{\celsius}.}} {We conclude that a larger change in temperature causes a larger change in \gls{hcfirst}.} 

\take{{RowHammer vulnerability (i.e., both \gls{ber} and \gls{hcfirst})} {tend} to worsen {as DRAM} temperature increases. However, individual DRAM rows can exhibit behavior {different} {from} this dominant trend.}

\subsection{{Circuit-level Justification}}
\label{sec:temperature_circuit}
We hypothesize that {our observations on} {the relation between RowHammer vulnerability and temperature } {are} caused by the non-monotonic behavior of charge trapping characteristics of DRAM cells.
{Yang et al.~\cite{yang2019trap} show a DRAM charge trap model simulated using a 3D~TCAD tool ({\emph{without}} real DRAM chip experiments).}
{The model} shows that  \gls{hcfirst} decreases as temperature increases, until a {temperature inflection} point where \gls{hcfirst} {starts} to increase as temperature increases.
{According to this model, a cell is more vulnerable to RowHammer at temperatures close to its temperature inflection point. 
We hypothesize that rows within a DRAM chip might have a {wide} variety of temperature {inflection} points, {and} thus the average temperature inflection point of a DRAM chip would determine whether the {average} RowHammer vulnerability increases or decreases with temperature
{(\obsrefs{temp:bounded}--\ref{temp:hcfirst_vs_temperature_larger})}.}
Park et al.~\cite{park2014active, park2016experiments} also {show} {an analysis of the relation between }\gls{hcfirst} {and DRAM temperature. Their observations are similar to ours,} but they consider {only} a small number 
of DDR3 DRAM cells.

Unlike simulations and {limited} results {reported} by~\cite{yang2019trap,park2014active, park2016experiments}, our {comprehensive} experiments with {\param{{272}}} DRAM chips show that 
{the temperature inflection points {for RowHammer vulnerability}  are very diverse {across DRAM cells and chips}}. 

%% file: 06_temporal.tex
\section{Aggressor Row Active Time Analysis}
\label{sec:aggressor_active_time}
{We} provide the first rigorous characterization of RowHammer considering the time that the aggressor row stays in the row buffer (i.e., \emph{{aggressor row active time}}).
{Prior {works}~\cite{park2014active, park2016experiments, walker2021ondramrowhammer} propose} circuit models and {suggest that RowHammer vulnerability of a victim row can depend on the aggressor row active time based on} preliminary data on a {very} small number of DRAM cells (i.e., only one carefully-selected DRAM row from each manufacturer)~\cite{park2014active, park2016experiments}.
However, none of these {works} conduct a rigorous analysis {of how} RowHammer vulnerability {varies} with aggressor row active time across a significant population of DRAM rows from real off-the-shelf DRAM modules. 

\begin{figure}[ht] 
    \centering
    \includegraphics[width=\linewidth]{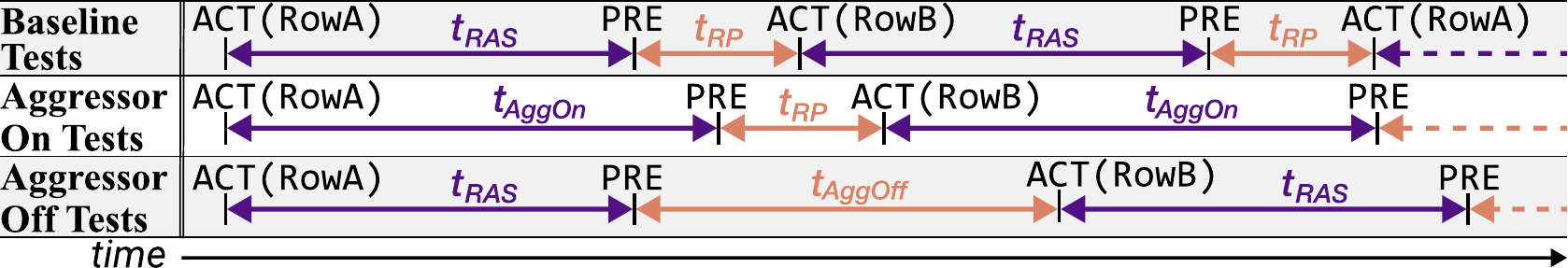}
    \vshort{}
    \vshort{}
    \caption{DRAM command timings for aggressor row active time (\gls{taggon}/\gls{taggoff})  experiments. Purple/Orange color indicates that an aggressor row is active/precharged.}
    \label{fig:tagg_on_off_procedure}
    \vspace{-3pt}
    
\end{figure}

{\figref{fig:tagg_on_off_procedure} describes the three test{s} we {perform} in our experiments:}  
{1) {\emph{Baseline Test}}, where we use $\tras{}$ as \gls{taggon}, and we use $\trp{}$ as \gls{taggoff},} 2)~\emph{Aggressor On Tests}, where we increase \gls{taggon} before the row is precharged (compared to {$\tras{}$ in Baseline Test}), and {3)}~\emph{Aggressor Off Tests}, where we increase \gls{taggoff} {before the aggressor row {is activated} (compared {to $\trp{}$ in Baseline Test})}.
Therefore, {for a given hammer count $HC$,} the overall attack time is $(t_{AggOn} + t_{RP}) \times HC$ and $(t_{RAS} + t_{AggOff}) \times HC$ for Aggressor On and Off {T}ests, respectively, while it is $(t_{RAS} + t_{RP}) \times HC$ for the baseline tests.
{Our experiments in this section are conducted at \SI{50}{\celsius} on the first 1K rows, the last 1K rows, and the 1K rows in the middle of a bank in our DDR4 chips.}

\label{sec:temporal}
\subsection{Impact of Aggressor Row's {On-}Time}
\begin{figure}[t] 
    \centering
    \includegraphics[width=\linewidth]{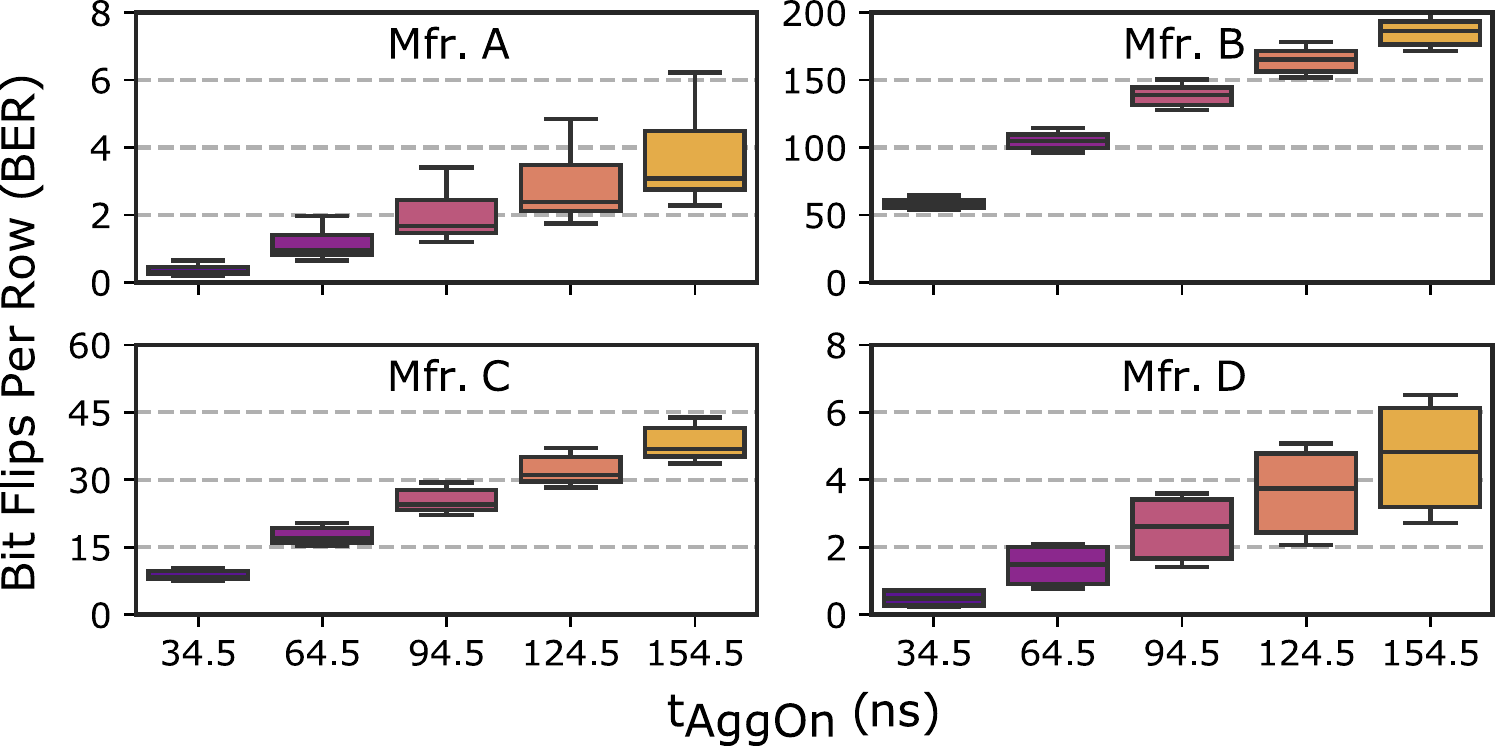}
    \caption{Distribution of the average number of bit flips per victim row {across chips as aggressor row} on-time (\gls{taggon}) {increases}.}
    \label{fig:ber_vs_tagg_on}
\end{figure}

\begin{figure}[h] 
    \centering
    \includegraphics[width=\linewidth]{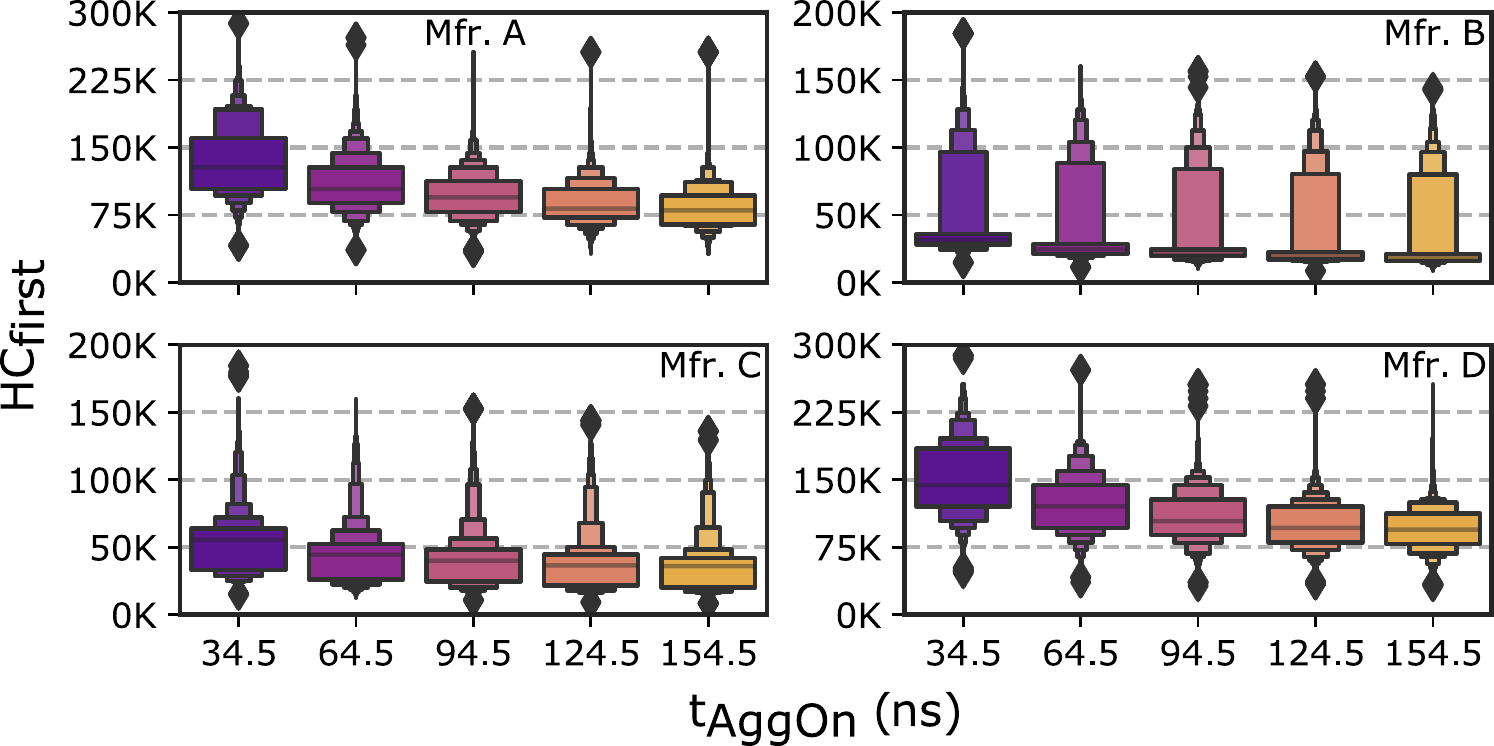}
    \caption{Distribution of {per-row} \gls{hcfirst} {across chips as aggressor row} on-time (\gls{taggon}) {increases}.}
    \label{fig:hcf_vs_tagg_on}
    \vshort{}
\end{figure}

{\figref{fig:ber_vs_tagg_on} and \figref{fig:hcf_vs_tagg_on} show the RowHammer bit flips per row (\gls{ber}) and \gls{hcfirst} distributions using box plots\footnote{{In a {box plot~\cite{Tukey1977Exploratory}}, {the} box shows the lower and upper quartile of the data (i.e., the box spans the $25^{\mathrm{th}}$ to the $75^{\mathrm{th}}$ percentile of the data). The line in the box represents the median. The bottom and top whiskers each represent {an} additional $1.5\times$ the \emph{inter-quartile range} (IQR, the range between the bottom and the top of the box) beyond the lower and upper quartile, respectively.}} and letter-value plots,\footnote{{In a letter-value plot~\cite{lvplot}, the widest box shows the lower and upper quartile of the data. The line in the box represents the median. The narrower box extended from the bottom of the widest box shows the lower octile ($12.5^{\mathrm{th}}$ percentile) and the lower quartile of the data, and the narrower box extended from the top of the widest box shows the upper octile and the upper quartile of the data, etc.. Boxes are plotted until all remaining data are outliers. Outliers are defined as the 0.7\% extreme values in the dataset, and are plotted as fliers in the plot.}} respectively, across all DRAM chips, as we vary \gls{taggon} {from \SI{34.5}{\nano\second} {($\tras{}$)} to \SI{154.5}{\nano\second}}.}

\observation{As the aggressor row stays active longer (i.e., \gls{taggon} increases), more DRAM cells experience RowHammer bit flips and they experience RowHammer bit flips at lower hammer counts.\label{taggon:vulnerability}}

{We observe that increasing \gls{taggon} from \SI{34.5}{\nano\second} to \SI{154.5}{\nano\second} \emph{{significantly}}  1)~}{increases \gls{ber} by $10.2\times,\ 3.1\times,\ 4.4\times,\ \text{and } 9.6\times$ on average}
{and 2)}~{decreases \gls{hcfirst}  by 40.0\%, 28.3\%, 32.7\%, and 37.3\% on average, in DRAM chips from Mfrs. A, B, C and D, respectively.}

\observation{RowHammer vulnerability consistently worsens as \gls{taggon} increases in DRAM chips from all four {manufacturers}.\label{taggon:bervariation}}

To see how  RowHammer vulnerability changes {as \gls{taggon} increases}, we examine \gls{cv}\footnote{$CV={standard~deviation/average}$~\cite{Everitt1998cambridgestatistics}.} values {of} {the} \gls{ber} and \gls{hcfirst} distributions {(not shown in the figures)}. We find that \gls{cv} {decreases} by around 15\% and 10\% for \gls{ber} and \gls{hcfirst}, respectively, across all four {manufacturers\om{,} as \gls{taggon} increases from \SI{34.5}{\nano\second} to \SI{154.5}{\nano\second}}. {This indicates that increasing {the} aggressor {row} active time {consistently worsens} RowHammer vulnerability across {the} DRAM chips {we test}.}

{{We conclude from }\obsrefs{taggon:vulnerability} and~\ref{taggon:bervariation} that increasing \gls{taggon} makes victim DRAM cells {much} more vulnerable to {a} RowHammer attack. {We exploit these observations in \secref{sec:implications}.}}

\take{{As {an aggressor row stays} active longer}, {victim DRAM cells become more vulnerable to RowHammer.}
\label{taggon:takeaway}}

\subsection{Impact of Aggressor Row's Off-Time}
\begin{figure}[h!] 
    \centering
    \includegraphics[width=\linewidth]{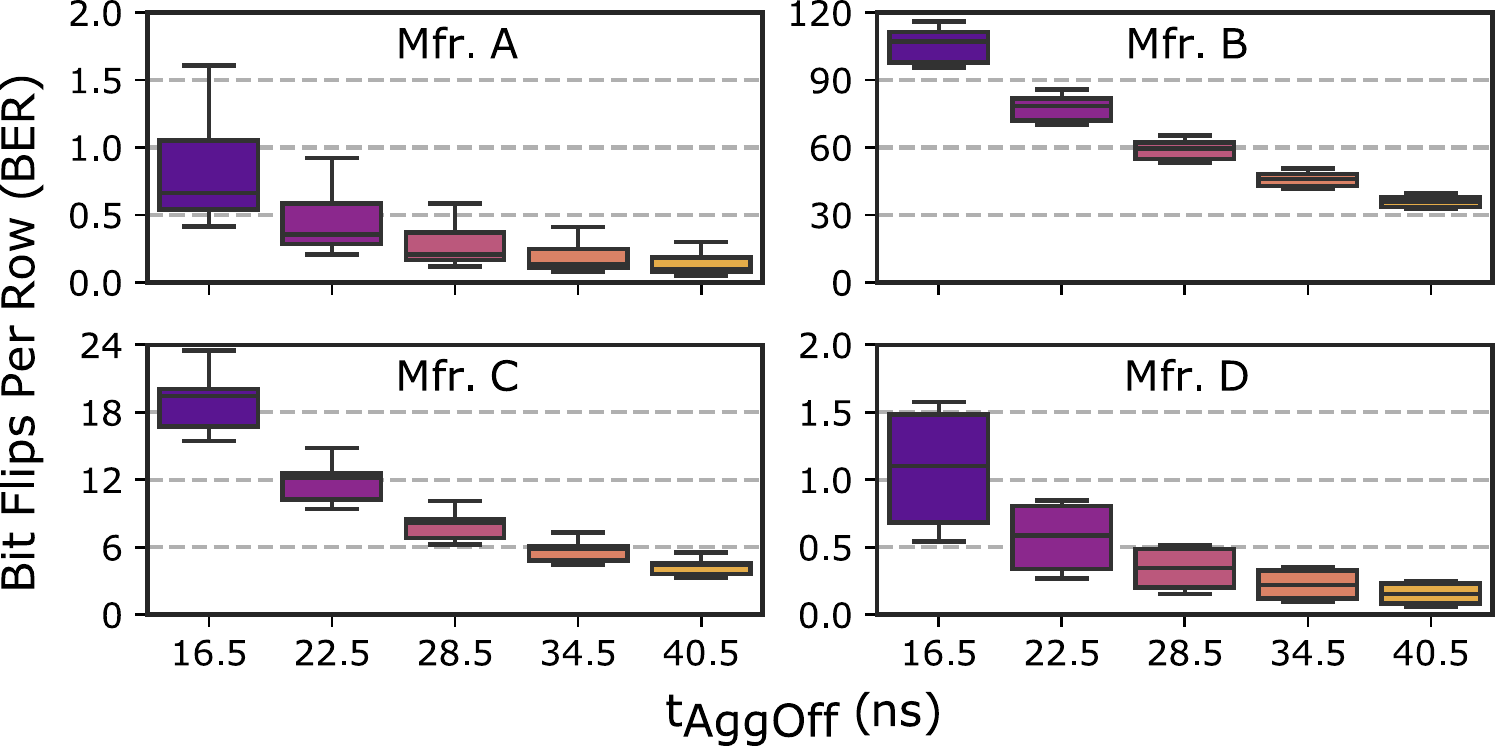}
    \caption{Distribution of the average number of {bit flips} per victim row {across chips as aggressor row} off-time (\gls{taggoff}) {increases}.}
    \label{fig:ber_vs_tagg_off}
\end{figure}

\begin{figure}[ht] 
    \centering
    \includegraphics[width=\linewidth]{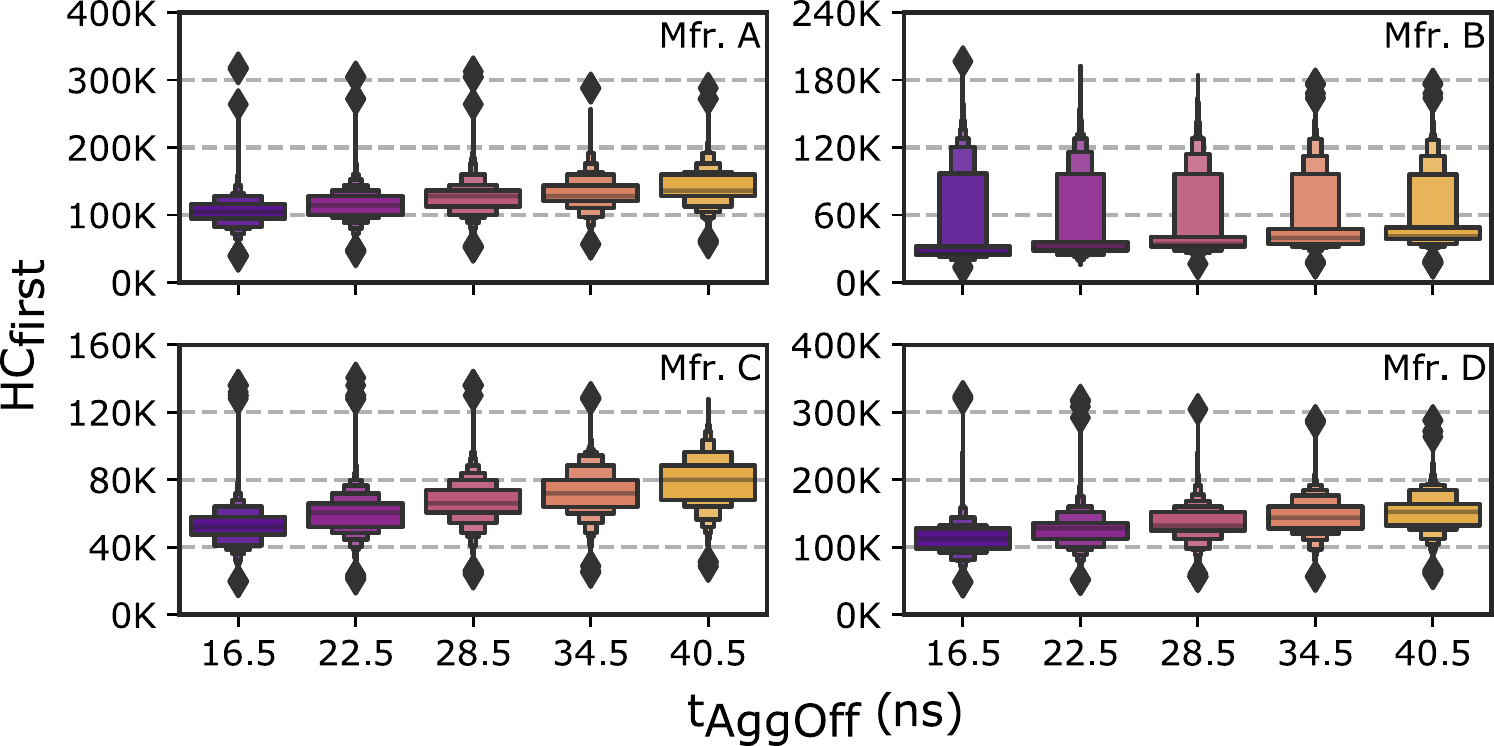}
    \caption{Distribution of {per-row} \gls{hcfirst} {across chips as aggressor row} off-time (\gls{taggoff}) {increases}.}
    \label{fig:hcf_vs_tagg_off}
    \vshort{}
\end{figure}

\figrefs{fig:ber_vs_tagg_off} and~\ref{fig:hcf_vs_tagg_off} show the \gls{ber} and \gls{hcfirst} distributions, {respectively, as we vary} \gls{taggoff} {from \SI{16.5}{\nano\second} {($\trp{}$)} to \SI{40.5}{\nano\second}}.{\footnote{Statistical configurations {of the box} and letter-value plots in \figrefs{fig:ber_vs_tagg_off} and~\ref{fig:hcf_vs_tagg_off} are identical {to those in} \figrefs{fig:ber_vs_tagg_on} and~\ref{fig:hcf_vs_tagg_on}, respectively.}}

\observation{
As the bank stays precharged longer (i.e., \gls{taggoff} increases), fewer DRAM cells experience RowHammer bit flips and they experience RowHammer bit flips at higher hammer counts.
\label{taggoff:vulnerability}}

{We observe that {increasing \gls{taggoff} {from \SI{16.5}{\nano\second} to \SI{40.5}{\nano\second}}} {\emph{significantly}} 1)~decreases \gls{ber} {by} $6.3\times,\ 2.9\times,\ 4.9\times,\text{ and } 5.0\times$ {on average,} and 2)~increases \gls{hcfirst} by 33.8\%, 24.7\%, 50.1\%, and 33.7\% {on average,} in DRAM chips from Mfrs. A, B, C, and D, respectively{.}
}

\observation{
RowHammer vulnerability consistently reduces {as \gls{taggoff} increases} in DRAM chips from all four manufacturers.\label{taggoff:bervariation}}

We observe that {the} \gls{cv} of \gls{hcfirst} {(not shown in the figures)} does not increase for any manufacturer
{as we increase }\gls{taggoff}. 
{Hence}, the level of reduction in RowHammer vulnerability is {similar} across {different rows'} \emph{most vulnerable cells}.
{In contrast,} {the} \gls{cv} of \gls{ber} increases by 18\% {on average} {for all} four manufacturers{, indicating {that the level} of reduction in RowHammer vulnerability {is different} across different rows}. 

{We conclude from} \obsrefs{taggoff:vulnerability} and~\ref{taggoff:bervariation} that increasing \gls{taggoff} makes it harder for a RowHammer {attack} to be successful. 
{We exploit this to improve} {RowHammer} {defense mechanisms} in \secref{sec:implications_defense}.

\take{{RowHammer vulnerability of victim cells decreases} when the {bank is precharged for a longer time}.}

\subsection{{Circuit-level Justification}}
\label{sec:temporal_circuit}
{{Prior work explains two circuit- and device-level mechanisms, causing RowHammer bit flips:} 
1) electron injection into the victim cell
~\cite{walker2021ondramrowhammer, yang2016suppression}, and 2) {wordline{-to}-wordline} cross-talk noise between {aggressor and victim rows that occurs} when the aggressor row is being activated~\cite{walker2021ondramrowhammer, ryu2017overcoming}. 
{We hypothesize that increasing the aggressor row's active time (\gls{taggon}) has a larger impact on exacerbating electron injection {to} the victim cell, 
compared to the reduction {in} cross-talk noise due to
lower activation frequency. Thus, RowHammer vulnerability worsens when \gls{taggon} {increases,} as our \obsrefs{taggon:vulnerability} and~\ref{taggon:bervariation} show.}
}

{{On the other hand, }{increasing a bank's precharged time (\gls{taggoff})}
{decreases} RowHammer vulnerability (\obsrefs{taggoff:vulnerability} and~\ref{taggoff:bervariation})
because longer \gls{taggoff} reduces the effect of cross-talk noise {without affecting electron injection} (since \gls{taggon} is unchanged). We leave the detailed device-level analysis and explanation of our observations to future works.}

%% file: 07_spatial.tex
\section{{Spatial Variation Analysis}}
\label{sec:spatial}

{We} provide the first {rigorous} {spatial variation} analysis of RowHammer across DRAM rows, subarrays, and columns. Prior work~\cite{kim2014flipping, kim2020revisiting, park2014active, park2016experiments, park2016statistical} analyzes RowHammer vulnerability {at the} DRAM bank granularity across many DRAM modules without {providing analysis of} the variation of this vulnerability across rows, subarrays, and columns. We provide this analysis and show that it is useful for improving both attacks and defense mechanisms.
{Our experiments in this section are conducted at \SI{75}{\celsius}.}

\subsection{Variation Across DRAM Rows} 
\label{sec:spatial_acrossrows}
\begin{figure}[h!] 
    \centering
    \includegraphics[width=\linewidth]{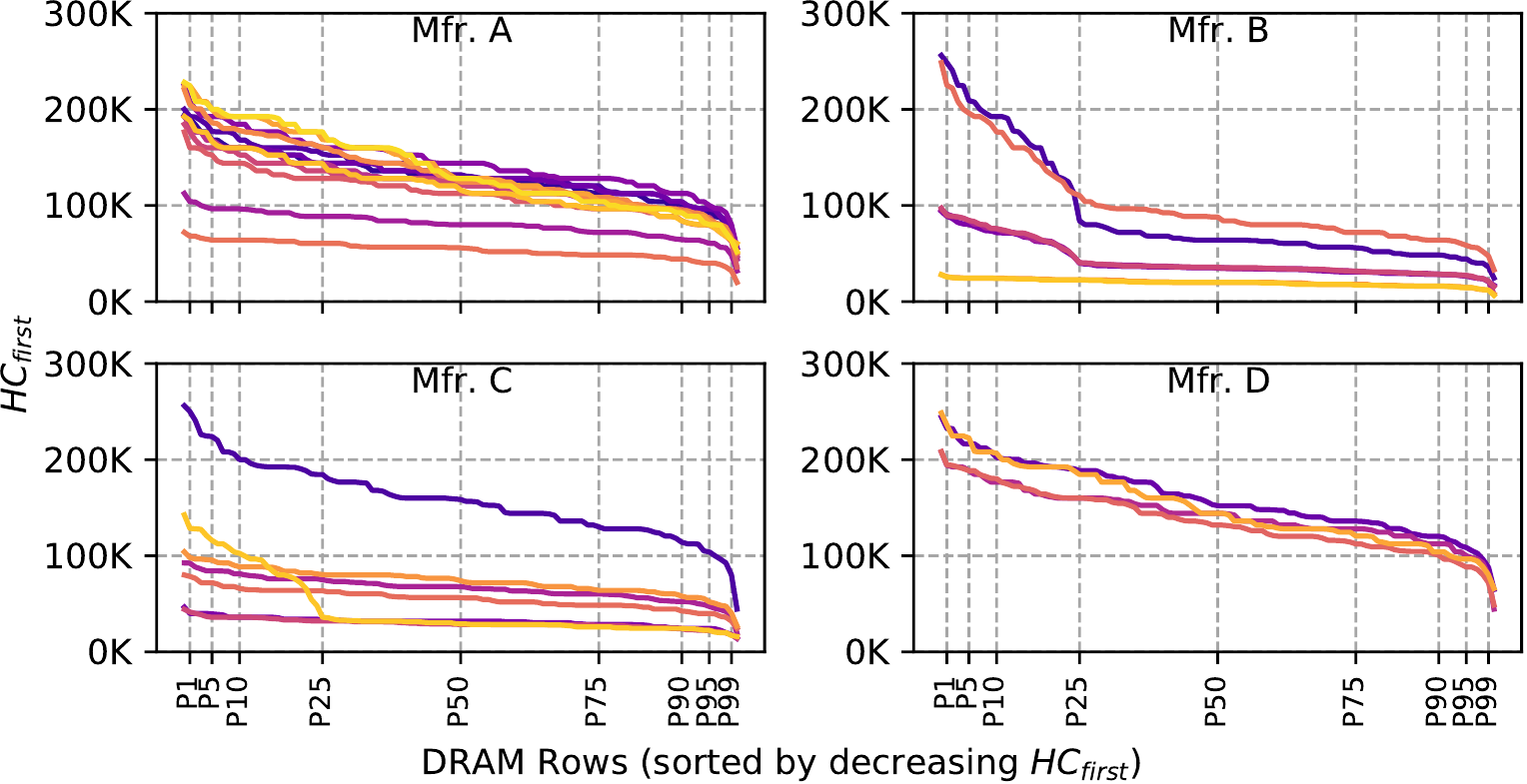}
    \vshort{}
    \vshort{}
    \caption{Distribution of \gls{hcfirst} across {vulnerable} DRAM rows. {Each curve represents a different tested DRAM module}.}
    \label{fig:hcfirst_across_rows}
\end{figure}

\figref{fig:hcfirst_across_rows} shows the {distribution of} \gls{hcfirst} {values} across {all vulnerable} DRAM rows {among the rows we test (\secref{sec:testing_methodology}). {For each row, we plot the minimum \gls{hcfirst} value observed across 5 repetitions of the test.} Each subplot {shows} DRAM modules from a different manufacturer, and each {curve} {corresponds to} a different DRAM module.}
The x-axis {shows} all the tested rows, sorted by decreasing \gls{hcfirst} {and marked with percentiles ranging from P1 to P99}.

\observation{A small fraction of DRAM rows are significantly more vulnerable to RowHammer than the vast majority of the rows.\label{spatial:hcfirst_across_rows}}

\gls{hcfirst} {varies significantly} across {rows}. We observe that 99\%, 95\%, and 90\% {of tested rows} {exhibit \gls{hcfirst} values {that are at least}}
\param{1.6$\times$}, \param{2.0$\times$}, and \param{2.2$\times$} {greater than} the {most vulnerable row's} \gls{hcfirst}{,} on average across all \param{four} {manufacturers}.
For example, 
{the lowest \gls{hcfirst} across all tested rows in a DRAM module from Mfr.~B is \param{33K},}
while 99\%, 95\%, and 90\% of the rows in the {same} {module} 
{exhibit \gls{hcfirst} values {equal to or greater than}}
\param{48.5K, 60.5K, and 64K}, respectively. {Therefore, we conclude that a small fraction of DRAM rows are significantly more vulnerable to RowHammer than the vast majority of the rows.}

The {large} variation {in} \gls{hcfirst} across DRAM rows can 
enable future improvements in
low-cost RowHammer {defenses} 
{(\secref{sec:implications_defense})}. 

\subsection{Variation Across Columns} 
\label{sec:spatial_acrosscols}
\begin{figure}[h!] 
    \centering
    \includegraphics[width=1.0\linewidth]{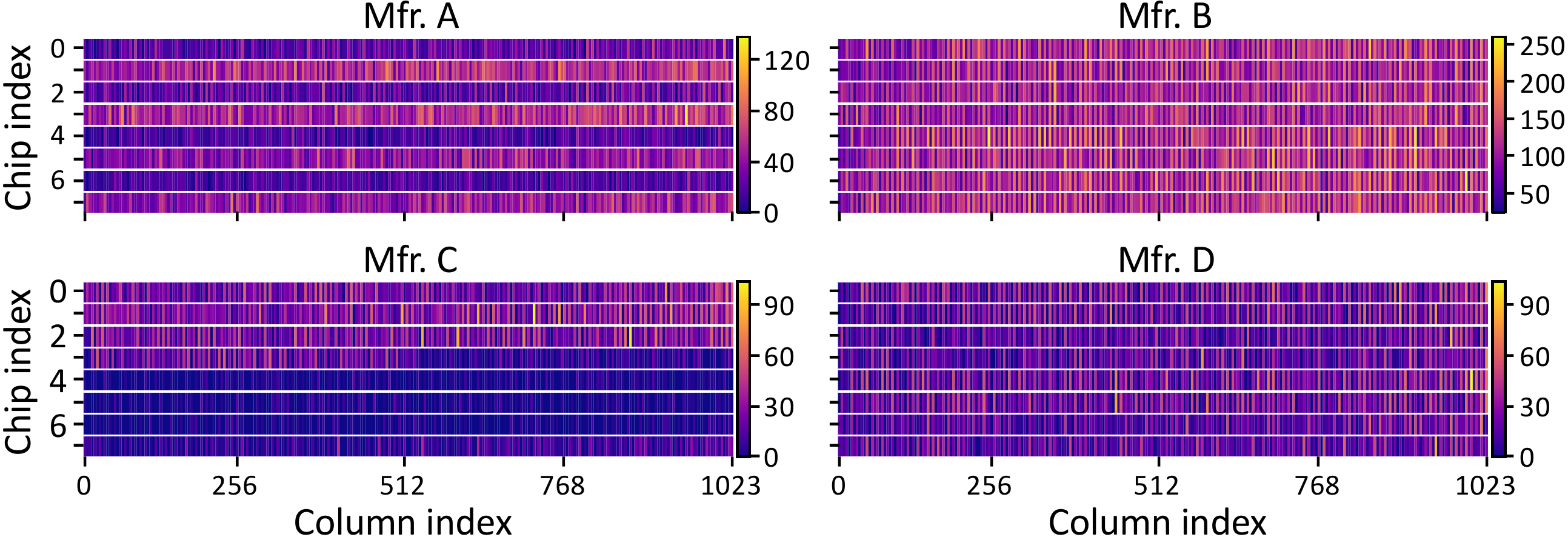}
      \vshort{}
      \vshort{}
      \vshort{}
    \caption{RowHammer bit flip distribution across columns in representative DRAM chips from four different manufacturers.}
    \label{fig:ber_across_columns}
    \vshort{}
\end{figure}

\figref{fig:ber_across_columns} shows
the distribution of {the number of} RowHammer bit flips across columns in {eight} representative DRAM {chip}s from {each of} \param{all four} manufacturers.
For each DRAM chip {(y-axis)}, we count the bit flips in each column {(x-axis)} across all {24K tested} rows.
The color-scale next to each subplot shows the bit flip count: a brighter color {indicates} more bit flips.

\observation{Certain columns are significantly more vulnerable to RowHammer than other columns.\label{spatial:columns}}

All chips show {significant variation} {in \gls{ber}} across columns. 
For example, the difference between the maximum and the minimum bit flip counts {per column} is larger than {\param{100} in modules from all four manufacturers.}
{Except for the module from Mfr.~B, where every column shows at least 6 bit flips, all the other tested modules have a considerable fraction of columns where \emph{no} bit flip occurs (27.80\%/31.10\%/9.96\% in Mfr.~A/C/D),
along with a very small faction of columns with more than 100 bit flips
(0.59\%/0.01\%/0.61\% in Mfr.~A/C/D).} {Therefore, we conclude that certain columns are significantly more vulnerable to RowHammer than other columns.}

\begin{figure}[ht!] 
    \centering
    \includegraphics[width=\linewidth]{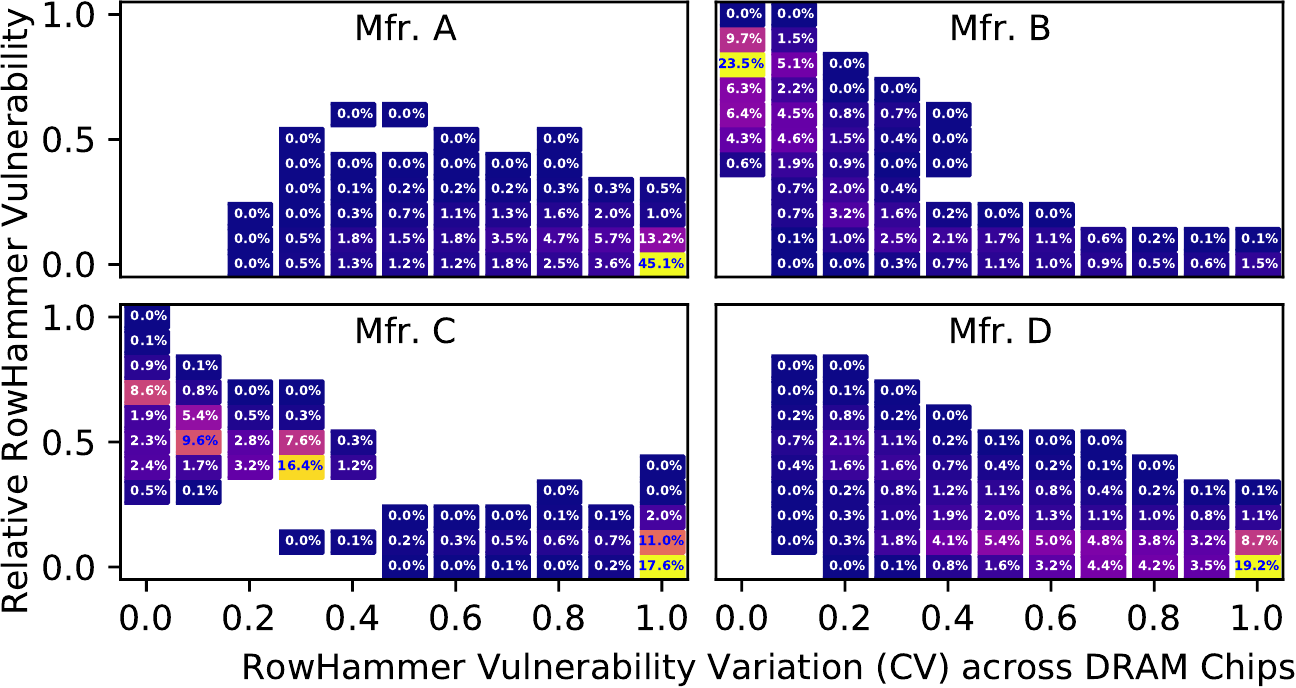}
    \vshort{}
    \vshort{}
    \caption{{Population of DRAM columns, clustered by relative RowHammer vulnerability. }
    }
    \label{fig:colvulnerability_across_chips}
    \vshort{}
\end{figure}

{To better understand this column-to-column variation, we study how RowHammer vulnerability varies between columns \emph{within} a single DRAM chip and \emph{across} different DRAM chips.}
{{Understanding this variation can provide insights into the {impact of circuit design} on a column's RowHammer vulnerability, {which is {important} for understanding and overcoming RowHammer}. A smaller variation in a column's RowHammer vulnerability across chips indicates a stronger influence of design-induced variation~\cite{lee2017design, kim2018solar}, while a larger variation across {chips that implement} the same design indicates a stronger influence of manufacturing process variation~\cite{lee2015adaptive, chang2016understanding, chang2017understanding, liu2012raidr, patel2017reaper, liu2013experimental,kim2019d,kim2018dram}.} {{To differentiate between these two sources of variation in our experiments}, we cluster {every column in a given DRAM module}
based on
two metrics.
The first metric is 
{the} column's \emph{relative RowHammer vulnerability}, defined as the column's \gls{ber}, normalized to the maximum \gls{ber} across all columns in the same module.
The second metric is \emph{the RowHammer vulnerability variation} at a column address.
We quantify the variation using 
the coefficient of variation (\gls{cv}) of the relative RowHammer vulnerability in columns with the same column address from different DRAM chips.}} \figref{fig:colvulnerability_across_chips} {shows a two-dimensional histogram with the \emph{relative RowHammer vulnerability} (y-axis) and \emph{Rowhammer vulnerability variation} (x-axis) uniformly quantized into 11 buckets each (i.e., 121 total buckets across each subplot).\footnote{We plot the x-axis as saturated at 1.0 because a \gls{cv} $>$ 1 means that the standard deviation is larger than the average, i.e., the variation is very large across chips.} Each bucket is illustrated as a rectangle containing a percentage value, which shows the percent of all columns that fall within the bucket. Empty buckets are omitted for clarity.}

\observation{Both design and manufacturing {processes} may affect {a DRAM column's RowHammer vulnerability.}\label{spatial:design_and_process}}

{We find that} {\param{50.9\%}} and {\param{16.6\%}}\footnote{{These numbers represent the population of columns whose \gls{cv} across chips is zero, i.e., sum of all annotated percentage values where \gls{cv}=0.}} of all vulnerable columns in DRAM modules from {Mfrs. B and C have \gls{cv}=0.0, which indicates that {each of these columns} {exhibit} the same level of RowHammer vulnerability {consistently} across {\emph{all}} DRAM chips {in a module}.} This {consistency} across chips
implies {that \emph{systematic variation} is {present}}, induced by {a chip's design{~\cite{lee2017design, kim2018solar,lee2013tiered,son2013reducing,vogelsang2010understanding,chang2016low,lee2015adaptive,chang2016understanding}}}.
{In contrast, \param{59.8\%, 30.6\%, and 29.1\%} of vulnerable columns in DRAM modules from {Mfrs.} A, C, and D show a {very} large variation across chips 
(\gls{cv}=1.0).}
This large variation 
across chips suggests {that
\emph{manufacturing process} variation is {\emph{also} a significant factor in} determining} a given DRAM column's RowHammer vulnerability.

{We conclude from} \obsrefs{spatial:hcfirst_across_rows}--\ref{spatial:design_and_process} that there is significant {variation} {in} RowHammer vulnerability across DRAM rows, columns, and chips. These observations {are} useful for 1)~crafting attacks that target vulnerable {locations} ({see} \secref{sec:implications_attack}) or 2)~improving defense mechanisms and error correction schemes that exploit the heterogeneity {of vulnerability} across DRAM rows and columns {({see} \secref{sec:implications_defense}).}

\take{RowHammer vulnerability significantly varies across DRAM rows and columns due to both design{-induced} and manufacturing{-}process{-induced} variation.}

\subsection{Variation Across Subarrays}
\label{sec:spatial_acrosssubarrays}
We analyze the RowHammer vulnerability of individual subarrays across DRAM chips. Since subarray boundaries are not publicly available, we {conservatively}
{assume a subarray size of 512~rows as reported in prior work~\cite{salp, lee2017design, chang2014improving, kim2018solar, vogelsang2010understanding}.{\footnote{{We verify this for some of our chips} {by performing 1)~single-sided RowHammer attack tests~\cite{kim2014flipping, kim2020revisiting} that induce bit flips in both rows adjacent to the aggressor row if the aggressor row is \emph{not} at the {edge} of a subarray and 2)~RowClone tests~\cite{seshadri2013rowclone, olgun2021quac, gao2019computedram} that can successfully copy data {only} between two rows within the same subarray.}}}}

\begin{figure}[t] 
    \centering
    \includegraphics[width=\linewidth]{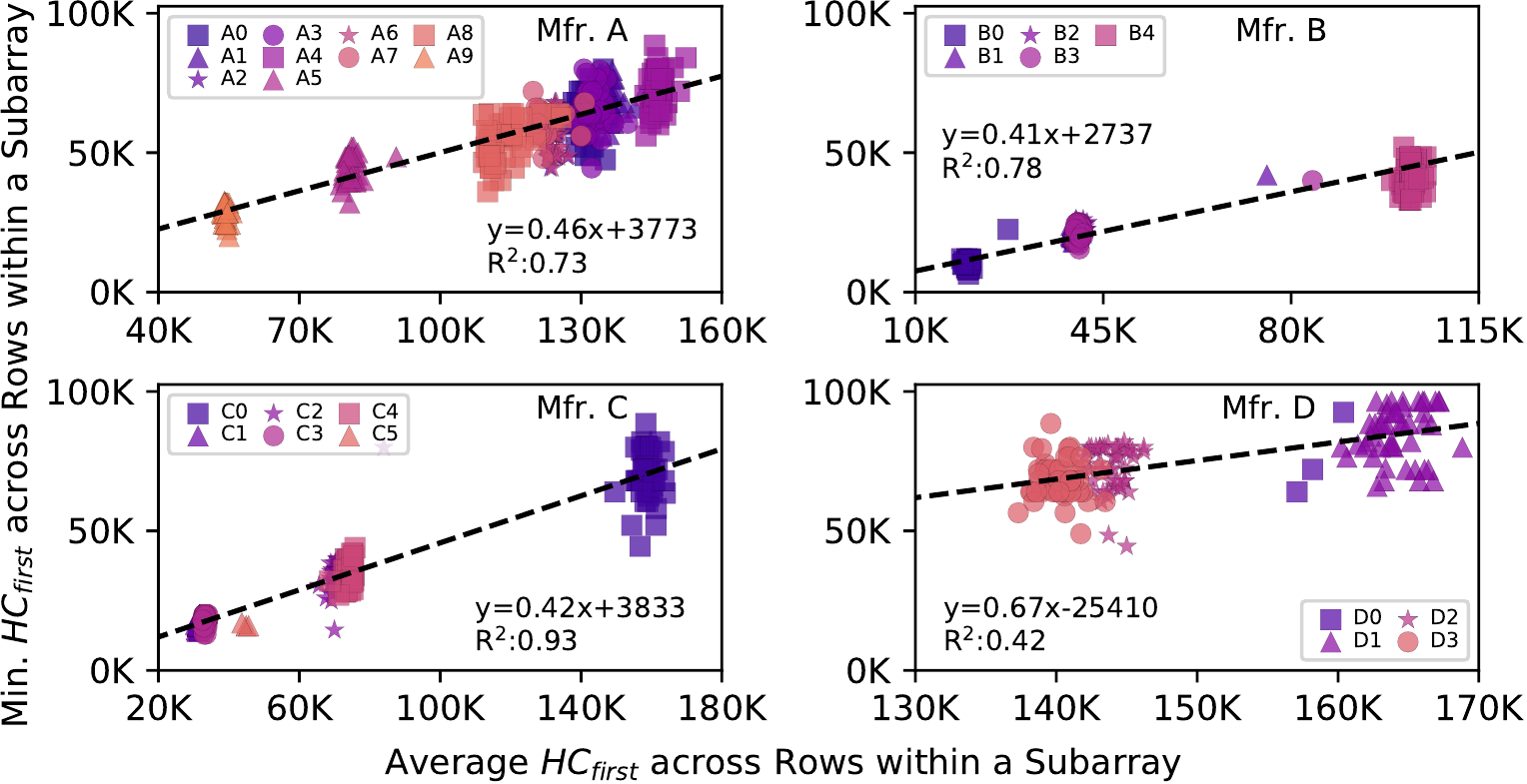}
    \vshort{}
    \vshort{}
    \caption{\gls{hcfirst} variation across subarrays. Each subarray is represented by the {average} (x-axis) and the {minimum} (y-axis) \gls{hcfirst} across the rows within the subarray.}
    \label{fig:hcfirst_across_subarrays}
    \vshort{}
    \vshort{}
\end{figure}

\figref{fig:hcfirst_across_subarrays} shows the variation of \gls{hcfirst} characteristics {in a DRAM bank} across subarrays {both 1) in a} DRAM module {and 2) across modules from the same manufacturer}. {Each color-marker pair represents a} different DRAM module. We represent the \gls{hcfirst} of a subarray in terms of {1)}~the {average} (x-axis) and {2)~the minimum} (y-axis) of \gls{hcfirst} across {the subarray's rows}. {For each manufacturer, }{we annotate a dashed line {that fits to the data via linear regression {with the specified $R^2$-score~\cite{wright1921correlation}.}}}

\observation{The most vulnerable DRAM row in a subarray is significantly more vulnerable than the other rows in the subarray.
\label{spatial:subarray_vs_bank}}

{We make two observations from \figref{fig:hcfirst_across_subarrays}.
First,
the average \gls{hcfirst} across all rows in a subarray is {on the order of 2$\times$} the most vulnerable row's \gls{hcfirst}, i.e., the minimum \gls{hcfirst}.
Therefore, the most vulnerable row in a subarray is \emph{significantly} more vulnerable than the other rows in the same subarray.
}
{{Second,} this relation between the minimum and average \gls{hcfirst} values is similar across subarrays from different modules from the same manufacturer, and thus can {be} {modeled using a linear regression}.
For example, {the minimum \gls{hcfirst} value in a subarray from Mfr.~C can be estimated using {a well-fitting} linear model with a $R^2$-score of 0.93.} 
{This observation is important because it indicates an underlying relationship between the average and minimum \gls{hcfirst} values across subarrays. For example, although subarrays in module~C0 have significantly larger \gls{hcfirst} values than subarrays from module~C3, a the linear model accurately expresses the relationship between both subarray's minimum and average \gls{hcfirst} values. Therefore, given {a} module from {Mfr.} C, the data shows that it may be possible to {predict} the minimum {(worst-case)} \gls{hcfirst} values of {another module's} subarrays, given {the} average {\gls{hcfirst}} values {of those subarrays.}}}

{We conclude from these two observations that 1)~the most vulnerable DRAM row in a subarray is significantly more vulnerable than the other rows in the subarray {and~2)~the}} worst-case \gls{hcfirst} in a subarray {can be predicted based on the average \gls{hcfirst} values and the linear models we provide.}

\begin{figure}[t] 
    \centering
    \includegraphics[width=\linewidth]{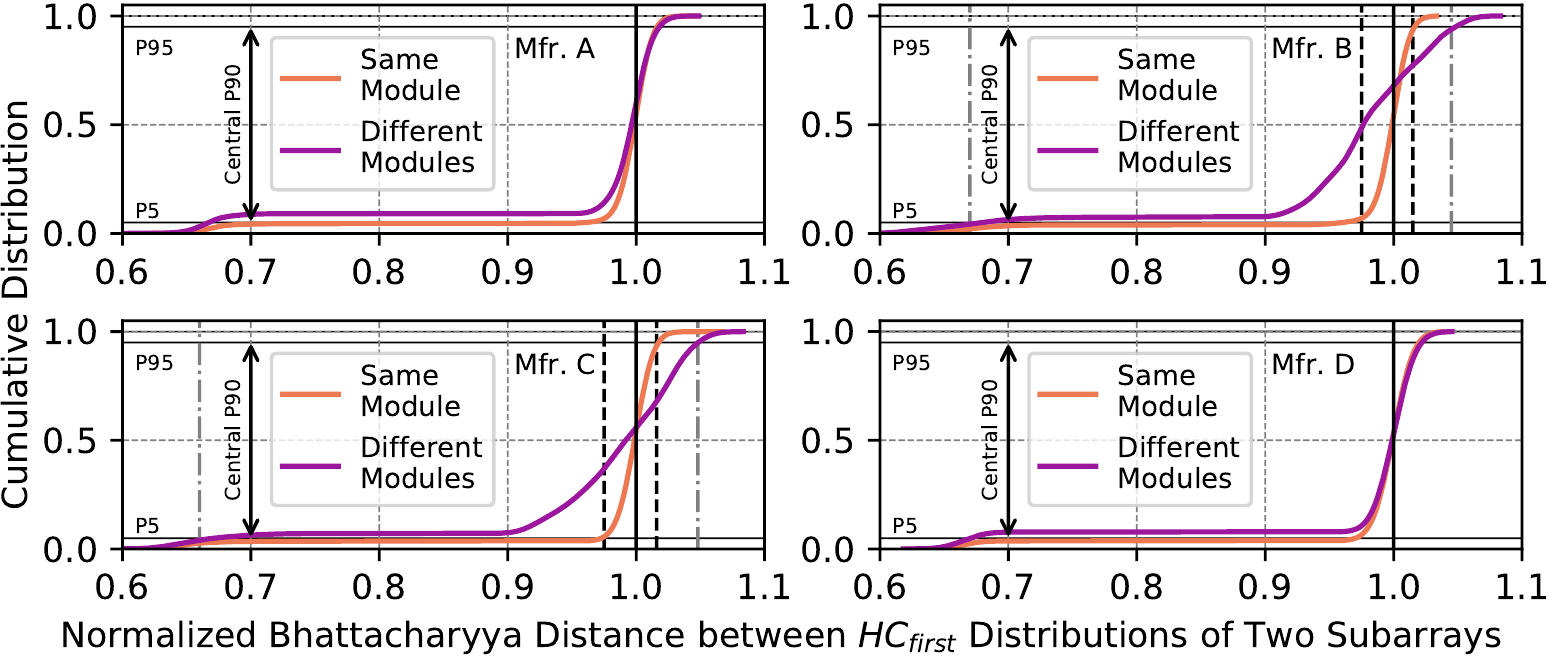}
    \vshort{}
    \vshort{}
    \caption{{Cumulative {distribution} of normalized Bhattacharyya distance {values between \gls{hcfirst} distributions {of} different subarrays} from 1) the same DRAM module and 2) different DRAM modules}}
    \label{fig:hcfirst_bd_across_subarrays}
    \vshort{}
\end{figure}

{To analyze and quantify the similarity between the RowHammer vulnerability of different subarrays, we statistically compare each subarray against all other subarrays from the same manufacturer. To compare two given subarrays, we first compare their \gls{hcfirst} distributions using Bhattacharyya distance ($BD$)~\cite{bhattacharyya1943measure}{,} which is used to measure the similarity of {two} statistical distributions. Second, for each pair of subarrays ($S_A$ and $S_B$), we normalize $BD$ to the $BD$ {between the first subarray $S_A$ and itself}: $BD_{norm} = BD(S_A, S_B) / BD(S_A, S_A)$. {Therefore}, $BD_{norm}$ is 1.0 if two distributions are identical, while $BD_{norm}$ value gets farther from 1.0 as the variation across two distributions increases.}
{\figref{fig:hcfirst_bd_across_subarrays} shows the cumulative distribution of $BD_{norm}$ values for subarray pairs from 1)~the same DRAM module and 2)~different DRAM modules. We annotate {P5, P95, and the central P90}
of the total population ({y-axis}) to show the range of $BD_{norm}$ values {in common-case.}}

\observation{\gls{hcfirst} distributions of subarrays within a DRAM module exhibit significantly more similarity to each other than \gls{hcfirst} distributions of subarrays from different modules.\label{spatial:bdist}}

{We observe that, when both $S_A$ and $S_B$ are from the same DRAM module (orange curves), the central 90th percentile {(i.e., between 5\% and 95\% of the population, as marked in~\figref{fig:hcfirst_bd_across_subarrays})}
of all subarray pairs exhibit $BD_{norm}$ values close to 1.0 (e.g., $BD_{norm}=0.975$ at the 5th {percentile} for Mfr. C), which means that their \gls{hcfirst} distributions are very similar. In contrast, $BD_{norm}$ values from different modules (purple curves) show a significantly wider distribution, especially for Mfrs. B and C (e.g., $BD_{norm}=0.66$ at the 5th {percentile} for Mfr. C).} 
{From} this {analysis,} {we conclude} that the \gls{hcfirst} 
{distribution within a subarray can be representative of other subarrays from the same DRAM module ({e.g., Mfrs.} B and C), {while} 
{the \gls{hcfirst} distribution within a subarray is often not representative of that of other subarrays for different DRAM modules.}}

\obsrefs{spatial:subarray_vs_bank} {and}~\ref{spatial:bdist} can be useful {for} improving DRAM profiling {techniques} and RowHammer {defense} mechanisms {(\secref{sec:implications_defense})}.

\take{{\gls{hcfirst} distribution in a subarray 1)~contains a diverse set of values and 2)~is similar to other subarrays in the same DRAM module.}}

\subsection{{{Circuit-level Justification}}}
\label{sec:spatial_circuit}
{We observe that RowHammer vulnerability {significantly} varies across DRAM rows, columns, {and chips, while different subarrays {in the same chip} exhibit similar vulnerability characteristics.} 

{\textbf{Variation across rows, columns, and chips.} 
We hypothesize that {two distinct factors cause the variation in RowHammer vulnerability that we observe across rows, columns, and chips.}}

{First, \emph{manufacturing process variation} causes differences in cell size and bitline/wordline impedance values, which introduces variation in} {cell} reliability characteristics {with}in and across {DRAM} chips~{\cite{lee2015adaptive, chang2016understanding, chang2017understanding, liu2012raidr, patel2017reaper, liu2013experimental,kim2019d,kim2018dram,orosa2021codic,olgun2021quac}}. {We hypothesize that similar imperfections in the manufacturing process (e.g., variation in cell-to-cell and cell-to-wordline spacings) cause RowHammer vulnerability to vary between cells in different DRAM chips.}

Second, {\emph{design-induced variation} causes cell access latency characteristics to vary deterministically based on a cell's physical location in the memory chip (e.g., its proximity to I/O circuitry)~\cite{kim2018solar, lee2017design}. In particular, prior work~\cite{lee2017design} shows that columns closer to wordline drivers (which are typically distributed along a row) can be accessed faster. Similarly, we hypothesize that columns that are closer to repeating analog circuit elements (e.g., wordline drivers, voltage boosters) more sensitive to RowHammer disturbance than columns that are farther away from such elements.}}

{\textbf{Similarity across subarrays.} Prior {works}~{\cite{lee2017design, kim2018solar}} demonstrate similar DRAM access latency characteristics across different subarrays. {This is because a cell's access latency is dominated by its physical distance from the peripheral structures (e.g., local sense amplifiers and wordline drivers) within the subarray{~\cite{chang2016low,lee2017design,lee2015adaptive,chang2016understanding,kim2018solar,lee2013tiered,son2013reducing,vogelsang2010understanding}}, causing {corresponding} cells in \emph{different} subarrays to exhibit \emph{similar} access latency characteristics.}
We hypothesize that different subarrays in a DRAM chip exhibit similar RowHammer vulnerability characteristics {for {a similar} reason}.} We leave {further analysis} {and validation} of these hypotheses for future work.

%% file: 08_implications.tex
\section{Implications}
\label{sec:implications}
The observations we make in \secref{sec:temperature}-\secref{sec:spatial} can be leveraged for both 1)~crafting more effective {RowHammer} attacks and 2)~{developing more effective and more efficient RowHammer defenses.}

\subsection{Potential Attack Improvements}
\label{sec:implications_attack}
Our {new observations and} characterization data can {help} improve the success probability of a RowHammer attack. We propose \param{three} attack improvements based {on} our {analyses of} temperature (\secref{sec:temperature}), {aggressor {row} active time (\secref{sec:aggressor_active_time}), and spatial variation (\secref{sec:spatial})}.

\textbf{Improvement 1.} {{\obsrefs{temp:bounded}--\ref{temp:narrow}}
can be used to craft more effective RowHammer attacks}
{where} the attacker can control or monitor the DRAM temperature. {{\obsrefs{temp:bounded}--\ref{temp:narrow}}} {show} that a DRAM cell is more vulnerable to RowHammer {within} a specific temperature range.
An attacker that can monitor the DRAM temperature (e.g., a malicious employee in a datacenter {or} an attacker who performs a remote RowHammer attack~\cite{lipp2018nethammer, tatar2018throwhammer} on a {physically accessible} IoT {device}) can increase the chance of a {bit flip} in two ways. First, the attacker can force the sensitive data to be {stored in} the DRAM cells that are more vulnerable at the {current} {operating} temperature, using known techniques~\cite{razavi2016flip,gruss2018another}. {Second, the attacker can heat up or cool down the chip to a temperature level at which the cells {that} stor{e} sensitive data become more vulnerable to RowHammer. 
As a result, the attacker can 
{significantly reduce the hammer count, {and consequently, the attack time, necessary} to cause a {bit flip}{, {thereby} reducing the probability of being detected}.} 
{For example, without our observations, an attacker {might} choose {an aggressor row based on an \emph{uninformed}} decision {with respect to} temperature characteristics.
In such {a} case, the chosen row could require a hammer count {larger than 100K (\figref{fig:hcfirst_across_rows}).}
However, by leveraging our {{\obsrefs{temp:bounded}--\ref{temp:narrow}}},
an attacker can make a {more \emph{informed} decision} and {choose a row whose \gls{hcfirst} reduces by 50\%}
(\figref{fig:hcfirst_variation})} at the temperature level the attack {is designed to} take place.}

\textbf{Improvement 2.} 
\obsref{temp:narrow} {can be used to} enable a new RowHammer attack {variant} as a temperature-dependent trigger of the main attack (which could be a RowHammer attack, or {some} other {security} attack).
{\obsref{temp:narrow}} demonstrates that some  DRAM  cells are  vulnerable to RowHammer in a very narrow temperature range.
{To implement a temperature-dependent trigger {using a RowHammer bit flip},} an attacker can place the victim data in a row that contains a cell that flips at the target temperature, 
{which allows the attacker to determine whether or not}
the target temperature is reached {to trigger the main attack}. 
{This could be useful for an attacker in two scenarios:
{1)~to trigger the attack only when a precise temperature is reached (e.g., triggering an attack against an IoT device in the field when the device is heated or cooled), and~2)~to identify abnormal operating conditions (e.g., triggering the attack {during peak hours} by using cells whose vulnerable temperature ranges are above the common DRAM chip temperature). For example, to detect that the temperature of a DRAM chip is precisely \SI{60}{\celsius} (above \SI{60}{\celsius}) an attacker can use the cells with a vulnerable temperature range of \SI{60}{\celsius}--\SI{60}{\celsius} ({all ranges with lower limit equal or higher than \SI{60}{\celsius}}), which are 0.3\%/0.3\%/0.3\%/0.2\% {(90.7\%/86.3\%/91.4\%/91.7\%)} of all vulnerable cells in Mfrs. A/B/C/D (\figref{fig:tempIntervals}).}}

\textbf{Improvement 3.} \obsref{taggon:vulnerability} {shows that} keeping an aggressor row active for a longer time results in more {bit flips} and lower \gls{hcfirst} values, {which can be used to craft more powerful RowHammer attacks.} 
{{For example, an} attacker} can increase the aggressor row active time by issuing more READ commands {to the aggressor row}{, which}
can potentially 1) increase the number of {bit flips} for a given hammer count, or 2) defeat already-deployed RowHammer defenses~\cite{yaglikci2021blockhammer,park2020graphene,lee2019twice, you2019mrloc, son2017making, seyedzadeh2018cbt, yaglikci2021security, devaux2021method, aweke2016anvil} {by inducing bit flips at a smaller hammer count than the \gls{hcfirst} value used for configuring a defense mechanism{.}}
{For example, issuing $10$~to~$15$ READ commands per aggressor row activation can increase the aggressor row active time by about $5\times$, increasing \gls{ber} by {3.2$\times$--10.2$\times$}
or causing bits to flip at a hammer count {that is} 36\% smaller than the \gls{hcfirst} value {that} {may be} used {to} configur{e} a defense mechanism {that does not consider} our {O}bservation{~\ref{taggon:vulnerability}}.}

\subsection{Potential Defense Improvements}
\label{sec:implications_defense}
Our characterization data can potentially be used in \param{five} ways to improve RowHammer defense methods. 

\textbf{Improvement 1.}
{\obsref{spatial:hcfirst_across_rows} {shows that there is a large spatial variation {in} \gls{hcfirst} across rows. A system designer} can {leverage this observation} to make existing RowHammer defense mechanisms more effective and efficient.}
{{A limitation of t}hese mechanisms {is that they are} configured for the smallest {\emph{(worst-case)}} \gls{hcfirst} across all rows in a DRAM bank {even though} {an overwhelming majority} of rows {exhibit significantly larger \gls{hcfirst} values}. This is an important limitation because, {when configured for a smaller \gls{hcfirst} value,} {the performance, {energy,} and area overheads of} {many RowHammer defense} mechanisms significantly {increase~\cite{kim2020revisiting, yaglikci2021blockhammer, park2020graphene}.} 
To {overcome this limitation,} a system designer can configure a RowHammer defense mechanism {to use} different \gls{hcfirst} values for different {DRAM rows}.}
For example, BlockHammer's~\cite{yaglikci2021blockhammer} and Graphene's~\cite{park2020graphene} area costs can reach approximately 0.6\% and 0.5\% of a high-end processor's die area~\cite{yaglikci2021blockhammer}. However, based on our \obsref{spatial:hcfirst_across_rows}, 95\% of DRAM rows {exhibit an \gls{hcfirst} value greater than $2\times$ the worst-case \gls{hcfirst}.}
Therefore, {both BlockHammer and Graphene
can be configured {with} {the worst-case \gls{hcfirst} for {only} 5\% of the rows and {with} $2\times$ \gls{hcfirst} for the 95\% of the rows,}
drastically reducing their area costs {down to 0.4\% and 0.1\% of the processor die area, translating to 33\% and 80\% area cost re{d}uction, respectively.}\footnote{{{Our preliminary evaluation} estimate{s} BlockHammer's~\cite{yaglikci2021blockhammer} and Graphene's~\cite{park2020graphene} area costs for $2\times$\gls{hcfirst}, following the methodology described in BlockHammer~\cite{yaglikci2021blockhammer}.}}
Similarly, the most {area-efficient} defense mechanism PARA~\cite{kim2014flipping} incurs 28\% slowdown on average {for} benign workloads when configured for an \gls{hcfirst} of 1K~\cite{kim2020revisiting}. {T}his {large} performance overhead can be halved~\cite{kim2020revisiting} for 95\% of the rows by simply using {lower} probability thresholds for {less vulnerable} rows. {We leave the comprehensive evaluation of such improvements to future work}.}

\textbf{Improvement 2.} \obsrefs{spatial:subarray_vs_bank} {and} \ref{spatial:bdist} on \emph{spatial {variation}} of \gls{hcfirst} {across subarrays} can be leveraged to {reduce the time required to profile a given DRAM module's RowHammer vulnerability characteristics.}
{{{This is an important challenge because p}rofiling a DRAM module's RowHammer characteristics {requires analyzing} several {environmental conditions and attack} properties (e.g., data pattern, access pattern, and temperature), {requiring {time-consuming tests that lead to long profiling times}}~\cite{kim2014flipping, kim2020revisiting, frigo2020trrespass, cojocar2020rowhammer, kwong2020rambleed, yang2019trap, park2016experiments, park2016statistical, patel2021harp}. 
According to our \obsrefs{spatial:subarray_vs_bank} {and} \ref{spatial:bdist}, characterizing a {\emph{small subset}} of subarrays can provide approximate {yet reliable} profiling data {for an {\emph{entire}} DRAM chip}. For example, assuming that a DRAM bank contains 128~subarrays, profiling {eight randomly-chosen} subarrays reduces RowHammer characterization {time} {by {at least} an order of magnitude. This low-cost approximate profiling can be useful in two cases.} 
First, finding the \gls{hcfirst} of a DRAM row requires performing a RowHammer {test} {with} {varying} hammer counts. Profiling the \gls{hcfirst} value for a few subarrays can be used {to limit} the \gls{hcfirst} search space for {the rows in the} rest of the subarrays based on our \obsref{spatial:bdist}. 
Second, one can profile a few subarrays within a DRAM module and use our linear regression models {(\obsref{spatial:bdist})} to estimate the DRAM module's RowHammer vulnerability for} systems whose reliability and security are not {as} critical (e.g., accelerators {and} {systems} running error-resilient workloads)~\cite{luo2014characterizing, koppula2019eden, nguyen2018approximate, nguyen2019stdrc, tu2018rana}.}

{\textbf{Improvement 3.} \obsrefs{temp:bounded} and~\ref{temp:narrow} show a vulnerable DRAM cell experiences bit flips at a particular temperature range. To improve a DRAM chip's reliability, {the system might incorporate a mechanism to temporarily or permanently retire DRAM rows (e.g., via software page offlining~\cite{meza2015revisiting} or hardware DRAM row remapping~\cite{yavits2020wolfram, carter1999impulse}) that are vulnerable to RowHammer within a particular operating temperature {range}. To adapt to changes in temperature, the row retirement mechanism might dynamically adjust the rows that are {retired}, potentially leveraging previously-proposed techniques (e.g., Rowclone~\cite{seshadri2013rowclone}, LISA~\cite{chang2016low}, NoM~\cite{rezaei2020nom}, {FIGARO~\cite{wang2020figaro}}) to efficiently move data between these rows.}}

\textbf{Improvement 4.} {\obsref{temp:ber_vs_temp}} demonstrates that overall \gls{ber} significantly increases with temperature across modules from three of the four manufacturers. {To reduce the success probability of a RowHammer attack, a system designer can improve the} cooling infrastructure for systems that use {such} DRAM modules{. Doing so} can reduce the number of RowHammer {bit flips} in a DRAM row.  
{For example,} {when} temperature {drops} from \SI{90}{\celsius} to \SI{50}{\celsius}, {\gls{ber} reduces by 25\%} {on average across DRAM modules from Mfr.~A}.  (see \figref{fig:ber_temp}).

\textbf{Improvement 5.} \obsref{taggon:vulnerability} shows that keeping an aggressor row {active} for a longer time increases the probability of RowHammer {bit flips}.
Therefore, RowHammer defenses should take aggressor row active time into account.
{Unfortunately, monitoring {the} active time of all potential aggressor rows {throughout} an entire refresh window is not feasible for emerging lightweight on-DRAM-die RowHammer defense mechanisms~\cite{yaglikci2021security, bennett2021panopticon, devaux2021method,jedec2020lpddr5, jedec2020ddr5}{, because such monitoring would {require} substantial storage and logic {to track} all potential aggressor rows' active times.}
{To address this issue}, the memory controller can {be modified to limit or reduce} {the} active time{s of} all rows {by changes to memory} request scheduling algorithms and{/or row buffer} policies {(e.g., via mechanisms similar to{~\cite{rixner00, kim2010atlas, subramanian2016bliss, yaglikci2021blockhammer,goossens2013conservative,huan2006processor, mutlu2008parbs, subramanian2014bliss,mutlu2007stall,moscibroda2007memory,kaseridis2011minimalistic})}}. {In this way}, a RowHammer defense mechanism {or the memory controller can inherently} {keep under control} an aggressor row's active time. {This is an example of a} system-DRAM cooperative scheme, similar {to} {those recommended by} prior work{~\cite{mutlu2013memory, kim2020revisiting, orosa2021codic, kim2014flipping,mutlu2017rowhammer}}.}

\textbf{Improvement 6.} \obsrefs{spatial:columns} and~\ref{spatial:design_and_process} show that {RowHammer vulnerability exhibits significant design-induced variation across columns within a chip and manufacturing process-induced variation across chips} in a DRAM module{.} {To make error correction codes (ECC) more effective and efficient {at} correcting RowHammer bit flips,} a system designer can {1)}~design {ECC} schemes optimized for non-uniform bit error probability distribution{s across columns} {and 2)}~modify the chipkill {ECC} {mechanism}~\cite{dell1997white,locklear2000chipkill, jian2013adaptive} to reduce {a} system's {dependency {on}} the most vulnerable DRAM chip, {as proposed in a} {concurrent work, revisit{ing} ECC {for} RowHammer~\cite{qureshi2021rethinking}.}

%% file: 09_related.tex
\section{Related Work}
\label{sec:relatedWork}

{{This} is the first {work} {that} rigorously {and experimentally} analyzes how RowHammer vulnerability changes with three fundamental properties: 1)~DRAM chip temperature, 2)~aggressor row active time, and 3)~victim DRAM cell’s physical location. 

{W}e {divide} prior work {on RowHammer} into {four} {categories}: 1)~attacks,
2)~defenses, 
{3})~characterization of real DRAM chips,
{and 4})~circuit-level simulation-based studies.
{Two works~\cite{mutlu2017rowhammer, mutlu2019rowhammer} provide an overview of {the} RowHammer {literature}, and project the effect of increased RowHammer vulnerability in future DRAM chips and DRAM-based memory systems.} 

{\textbf{RowHammer Attacks and Defenses.} Many works~\cite{seaborn2015exploiting, van2016drammer, gruss2016rowhammer, razavi2016flip, pessl2016drama, xiao2016one, bosman2016dedup, bhattacharya2016curious, qiao2016new, jang2017sgx, aga2017good, mutlu2017rowhammer, tatar2018defeating, gruss2018another, lipp2018nethammer, van2018guardion, frigo2018grand, cojocar2019eccploit,  ji2019pinpoint, mutlu2019rowhammer, hong2019terminal, kwong2020rambleed, frigo2020trrespass, cojocar2020rowhammer, weissman2020jackhammer, zhang2020pthammer, rowhammergithub, yao2020deephammer, hassan2021utrr} exploit the RowHammer vulnerability to induce bit flips in main memory{,}
as \secref{sec:background_rowhammer} {explains}. These works activate two (double-sided attack~\cite{kim2014flipping, kim2020revisiting, seaborn2015exploiting})
or more (many-sided attack~\cite{frigo2020trrespass}) aggressor rows, \emph{as rapidly as possible}, aiming to maximize the number of RowHammer-induced bit flips. {However, these works} do \emph{{not}} consider RowHammer's sensitivities to temperature, {aggressor} {row} active time, and spatial variation.
Similarly, existing RowHammer defense mechanisms~\cite{AppleRefInc, kim2014flipping, kim2014architectural, aichinger2015ddr, bains2015row, aweke2016anvil, bains2016distributed, bains2016row, gomez2016dummy, brasser2017can, son2017making, konoth2018zebram, seyedzadeh2018cbt, van2018guardion, hassan2019crow, lee2019twice, kang2020cattwo, park2020graphene, yaglikci2021blockhammer, yaglikci2021security, devaux2021method, you2019mrloc, jedec2017ddr4, jedec2015hbm, jedec2020ddr5, jedec2020lpddr5}
are not designed to account for these three properties. 
The new {observations and} insights we provide
can be used to improve {both} RowHammer attacks and defenses, as \secref{sec:implications} {describes}. We leave {a} full exploration of such attacks and defenses to future work, as our goal is to develop a fundamental understanding of RowHammer properties as opposed to developing new attacks and defenses.} 
} 

\textbf{Characterization of Real DRAM Chips.} {Two major works} extensively characterize the RowHammer vulnerability {using} real DRAM chips~\cite{kim2014flipping, kim2020revisiting}. The {original RowHammer work}~\cite{kim2014flipping}, published in 2014, 1)~investigates the vulnerability of {129} commodity {DDR3} DRAM {modules} to various RowHammer attack models, 2)~{demonstrates for the first time} that RowHammer is a real problem for commodity DRAM chips, {3)~characterizes RowHammer's sensitivity to refresh rate and activation rate in terms of \gls{ber}, \gls{hcfirst}, and {the physical distance between aggressor and victim rows}}, and {4}) {examines various potential solutions and} proposes a {new} low-cost mitigation mechanism. 

{The second} work~\cite{kim2020revisiting}, published in 2020, conducts {comprehensive} {scaling} experiments on a {wide} range of {1580 DDR3, DDR4, and LPDDR4 commodity DRAM chips from different DRAM generations and technology nodes}, clearly {demonstrating} that RowHammer {has} {become {an even} more serious} problem {over {DRAM} generations.}
Even though these two works rigorously characterize {various aspects of the RowHammer vulnerability in} real DRAM chips, {they} do not analyze {the effects of temperature, aggressor row active time, and victim DRAM cell's physical location on the} RowHammer {vulnerability}.
Our work {complements and furthers the analys{e}s} o{f} these two papers~\cite{kim2014flipping, kim2020revisiting} {by} {1)~rigorously {analyzing} how these three properties affect the RowHammer vulnerability, and 2)~{providing} new {insights} {in}to craft{ing} more effective {and efficient} RowHammer attacks and  defenses.}

{Three other} works~\cite{park2014active, park2016experiments, park2016statistical} {present \emph{preliminary}} experimental {data from only three~\cite{park2014active, park2016statistical} or five~\cite{park2016experiments} DDR3 DRAM chips to build models {that explain} how {the RowHammer vulnerability of DRAM cell{s}} varies with the three properties we analyze. Unfortunately, the {experimental data {provided by} these works} is not rigorous and conclusive enough due to {1)}~their {extremely small} sample set {of DRAM cells, rows, and chips} and {2)}~the lack of analysis {of} system-level implications.}
{Our work, {in contrast,} 1)~\emph{rigorously} {analyzes} the effects of all three properties by {testing} a significantly larger set of {272} DRAM chips, and 2)~{provides insights into resulting}
RowHammer attack and defense improvements.}

\textbf{Simulation-based Studies.} {Prior works}~\cite{redeker2002investigation, ryu2017overcoming, yang2016suppression, yang2017scanning, yang2019trap, gautam2019row, jiang2021quantifying, walker2021ondramrowhammer} attempt {to} explain the error mechanisms {that cause} RowHammer {bit flips} {through circuit-level simulations of capacitative-coupling and charge-trapping mechanisms{, without testing real DRAM chips.} {These works, {some of which we discuss in \secref{sec:temperature_circuit} and~\secref{sec:temporal_circuit}}}{,} are orthogonal to our experimental study.}

%% file: 10_conclusion.tex
\section{Conclusion}
\label{sec:conclusion}
This work provides the first study that experimentally analyzes the impact of DRAM chip temperature, aggressor row active time, and victim DRAM cell's physical location on RowHammer vulnerability, through extensive characterization of {real DRAM chips}.
We rigorously characterize 248 DDR4 and 24 DDR3 modern DRAM chips {from four major DRAM manufacturers using} {a} carefully designed methodology and metrics, {providing} 16 key observations and 6 key takeaways. {W}e highlight {three major observations:} 1)~a DRAM cell experiences RowHammer bit flips at a bounded temperature range, 2)~a DRAM row is more vulnerable to RowHammer when the {aggressor row stays active for longer}, and 3)~a small fraction of DRAM rows are significantly more vulnerable to RowHammer than the other rows within {a} DRAM module.
{We {describe and analyze} how our insights can be used to improve both RowHammer attacks and defenses.}
We hope that the {novel experimental} results and insights of our study will inspire and aid future work to develop effective {and} efficient {solutions to the RowHammer problem.}

%% file: 11_appendix.tex
\section{Appendix}
\label{sec:appendix}
\newcommand*{\myalign}[2]{\multicolumn{1}{#1}{#2}}

Table~\ref{tab:detailed_info} shows the characteristics of the DDR4 and DDR3 DRAM modules we test and analyze.
\begin{table*}[ht]
\footnotesize
\centering
\caption{Characteristics of the tested DDR4 and DDR3 DRAM modules.}
\label{tab:detailed_info}
\begin{tabular}{l|l|l||l|l|c|c|c|c|c|c|c}
\textbf{Type}     & \myalign{c|}{\textbf{\begin{tabular}[c]{@{}c@{}}Chip \\ Manufacturer\end{tabular}}} & \myalign{c||}{\textbf{\centering\begin{tabular}[c]{@{}c@{}}Chip \\ Identifier\end{tabular}}} & \myalign{c|}{\textbf{\begin{tabular}[c]{@{}c@{}}Module \\ Vendor\end{tabular}}} & \myalign{c|}{\textbf{\begin{tabular}[c]{@{}c@{}}Module \\ Identifier\end{tabular}}}              & \textbf{\begin{tabular}[c]{@{}c@{}}Freq. \\ (MT/s)\end{tabular}} & \textbf{\begin{tabular}[c]{@{}c@{}}Date \\ Code\end{tabular}} & \textbf{Density}     & \textbf{\begin{tabular}[c]{@{}c@{}}Die \\ Rev.\end{tabular}} & \textbf{Org.}         & \textbf{\begin{tabular}[c]{@{}c@{}}\#Modules\end{tabular}} & \textbf{\begin{tabular}[c]{@{}c@{}}\#Chips\end{tabular}} \\ \hline
\hline
\multirow{6}{*}{DDR4} & \multirow{3}{*}{A: Micron}                                               & \multirow{3}{*}{MT40A2G4WE-083E:B}                                  & \multirow{3}{*}{Micron}                                           & \multirow{3}{*}{\begin{tabular}[c]{@{}c@{}}MTA18ASF2G72PZ-\\ 2G3B1QG~\cite{datasheetmta18asf2g72pz}\end{tabular}} & \multirow{3}{*}{2400}                                            & 1911                                                          & \multirow{3}{*}{8Gb} & \multirow{3}{*}{B}                                           & \multirow{3}{*}{x4} & 6  & 96                                                            \\ \cline{7-7} \cline{11-12}
                      &                                                                       &                                                                     &                                                                   &                                                                                    &                                                                  & 1843                                                          &                      &                                                              &                     & 2  &32                                                            \\ \cline{7-7} \cline{11-12}
                      &                                                                       &                                                                     &                                                                   &                                                                                    &                                                                  & 1844                                                          &                      &                                                              &                     & 1  &16                                                            \\ \cline{2-12} 
                      & B: Samsung                                                               & K4A4G085WF-BCTD~\cite{datasheetK4A4G085WF}                                                     & G.SKILL                                                           & F4-2400C17S-8GNT~\cite{datasheetf42400c17s}                                                                   & 2400                                                             & 2021 Jan $\star$                                                    & 4Gb                  & F                                                            & x8                  & 4 & 32                                                              \\ \cline{2-12} 
                      & C: SK Hynix                                                                 & DWCW (Partial Marking)      $\dag$                                         & G.SKILL                                                           & F4-2400C17S-8GNT~\cite{datasheetf42400c17s}                                                                   & 2400                                                             & 2042                                                      & 4Gb                  & B                                                            & x8                  & 5      & 40                                                        \\ \cline{2-12} 
                      & D: Nanya                                                                 & D1028AN9CPGRK $\ddag$                                                        & Kingston                                                          & KVR24N17S8/8~\cite{datasheetkvr24n17s8}                                                                       & 2400                                                             & 2046                                                          & 8Gb                  & C                                                           & x8                  & 4     &32                                                         \\ \hline
              \hline
\multirow{3}{*}{DDR3} & A: Micron                                                                & MT41K512M8DA-107:P~\cite{datasheetMT41K512M8DA}                                                  & Crucial                                                           & CT51264BF160BJ.M8FP                                                                & 1600                                                             & 1703                                                          & 4Gb                  & P                                                            & x8                  & 1             &8                                                 \\ \cline{2-12} 
                      & B: Samsung                                                               & K4B4G0846Q                                                          & Samsung                                                           & M471B5173QH0-YK0~\cite{datasheetM471B5173QH0}                                                                   & 1600                                                             & 1416                                                          & 4Gb                  & Q                                                            & x8                  & 1   &8                                                           \\ \cline{2-12} 
                      & C: SK Hynix                                                                 & H5TC4G83BFR-PBA                                                     & SK Hynix                                                             & HMT451S6BFR8A-PB~\cite{datasheetHMT451S6BFR8A}                                                                   & 1600                                                             & 1535                                                          & 4Gb                  & B                                                            & x8                  & 1  &8                                                           
\\ \hline \hline
\end{tabular}
\begin{flushleft}
$\star$ We use the date marked on the modules due to the lack of date information on the chips.

$\dag$ A part of the chip identifier is removed on these modules. We infer the DRAM chip manufacturer and die revision information based on the remaining part of the chip identifier.

$\ddag$ We extract the DRAM chip manufacturer and die revision information from the serial presence detect (SPD) registers on the modules.
\end{flushleft}
\end{table*}